\documentclass{article}
\usepackage{graphicx}
\usepackage{amsmath}
\usepackage{amsfonts}
\usepackage{amssymb, soul}
\usepackage{natbib}

\usepackage{graphicx}
\usepackage{epstopdf}
\usepackage{amssymb}
\usepackage{amsmath}
\usepackage[rightcaption]{sidecap}
\usepackage[export]{adjustbox}
\usepackage{float}
\usepackage{url}

\usepackage{epstopdf}

\newcommand{\be}{\begin{equation}}
\newcommand{\ee}{\end{equation}}

\begin{document}

\title{Chaotic dynamics in the planar gravitational many-body problem with rigid body rotations}

\author{James A. Kwiecinski, Attila Kovacs, Andrew L. Krause,\\
 Ferran Brosa Planella, Robert A. Van Gorder\\
Mathematical Institute, University of Oxford, Andrew Wiles Building\\
 Radcliffe Observatory Quarter, Woodstock Road, Oxford, OX2 6GG, United Kingdom\\
Robert.VanGorder@maths.ox.ac.uk}

\maketitle

\begin{abstract}
The discovery of Pluto's small moons in the last decade brought attention to the dynamics of the dwarf planet's satellites. With such systems in mind, we study a planar $N$-body system in which all the bodies are point masses, except for a single rigid body. We then present a reduced model consisting of a planar $N$-body problem with the rigid body treated as a 1D continuum (i.e. the body is treated as a rod with an arbitrary mass distribution). Such a model provides a good approximation to highly asymmetric geometries, such as the recently observed interstellar asteroid 'Oumuamua, but is also amenable to analysis. We analytically demonstrate the existence of homoclinic chaos in the case where one of the orbits is nearly circular by way of the Melnikov method, and give numerical evidence for chaos when the orbits are more complicated. We show that the extent of chaos in parameter space is strongly tied to the deviations from a purely circular orbit. These results suggest that chaos is ubiquitous in many-body problems when one or more of the rigid bodies exhibits non-spherical and highly asymmetric geometries. The excitation of chaotic rotations does not appear to require tidal dissipation, obliquity variation, or orbital resonance. Such dynamics give a possible explanation for routes to chaotic dynamics observed in $N$-body systems such as the Pluto system where some of the bodies are highly non-spherical.\\
\\
\textit{keywords}: celestial mechanics; many body problem; rigid bodies; chaotic dynamics; Melnikov method
\end{abstract}

\section{Introduction}
The study of $N$-body problems, whereby the bodies are treated as point masses, has a long and storied history \cite{musielak2014three}. There has been much effort on finding new orbits to the planar three-body problem, with a large number of new orbits found over the last few decades \cite{moore1993braids,chenciner2000remarkable,nauenberg2001periodic,chenciner2005rotating,nauenberg2007continuity,dmitravsinovic2015topological}. In particular, very recent work has uncovered hundreds of new orbits in the planar three-body problem \cite{vsuvakov2013three,li2017one,dmitravsinovic2017newtonian}. In all such studies, one assumes that each mass is a point mass. There is also work done on the statistical mechanics of $N$-body problems \cite{lynden1995landau,lynden1999exact} as well as extensions to the relativistic regime \cite{eddington1938problem}.

In contrast to the $N$-body problem for point masses, there have been relatively fewer studies on orbital dynamics when the masses are rigid bodies. There have been a number of recent studies on the rotation of rigid bodies \cite{lara2014short,hou2017mutual,boue2017two}. It was recently shown that a relative equilibrium motion of $N\geq 3$ disjoint rigid bodies is never an energy minimizer \cite{moeckel2017minimal}, a generalization of a known result about point masses in the $N\geq 3$ body problem \cite{moeckel1990central} to the case of rigid bodies.

Rigid body dynamics are of great relevance to applications in celestial mechanics when a mass is non-spherical. Small bodies with diameters smaller than 1000 km usually have irregular shapes, often resembling dumbbells or contact binaries \cite{melnikov2010rotation,jorda2016global,lages2017chaotic}. The spinning of such a gravitating dumbbell creates a zone of chaotic orbits around it \cite{mysen2006chaotic,frouard2012instability,mysen2009predictability}, and these dynamics were recently studied in \cite{lages2017chaotic}, where it was shown that the chaotic zone increases in size if the rotation rate is decreased. To do so, \cite{lages2017chaotic} generalized the Kepler map technique \cite{meiss1992symplectic} to describe the motion of a particle in the gravitational field of a such a rotating irregular body modeled by a dumbbell. Regarding the rotations of such an irregular object itself, \cite{fernandez2016dynamics} recently studied dynamics from planar oscillations of a dumbbell satellite orbiting an oblate body of much larger mass, obtaining a kind of generalized Beletsky equation. They show the existence of chaotic orbits in their system via the Melnikov method, and study the transition between regular to chaotic orbits numerically through the use of Poincar\'e maps. The model studied was originally derived in \cite{abouelmagd2015dynamics}, and while interesting, is fairly restrictive on the geometry and mass distribution in the rigid body, while further assuming that the satellite body is of much smaller mass to the oblate body it orbits.

One of the large astronomical bodies to first get attention for its possible chaotic motion was Hyperion, a highly non-spherical moon of Saturn. In \cite{wisdom1984chaotic}, the authors analyze the spin-orbit coupling of Hyperion, modeled as a homogeneous ellipsoid orbiting around a point mass, to determine the critical oblateness of the ellipsoid that induces chaos in the system. They conclude their analysis with a prediction of chaotic tumbling for Hyperion. Later, \cite{tarnopolski2017rotation} further studied the motion of oblate satellites and applied the results to Hyperion as well. A control term was introduced to the Hamiltonian of the system in order to reduce the chaotic behavior; numerical simulations showed that the control term manages to effectively reduce the chaotic behavior of the system and, for some parameter values, it even succeeds in suppressing chaos. On the other hand, \cite{tarnopolski2017influence} studied the effect of a second satellite in the dynamics of an oblate moon, applied in particular to the effects of Titan in the Saturn-Hyperion system. Building up from the model used in \cite{wisdom1984chaotic}, Tarnopolski introduces a second satellite as a point mass. The numerical simulations of the model show how the presence of a second satellite can result in a transition of periodic orbits into chaos. Instead of spin-orbit coupling, spin-spin coupling is presented in \cite{batygin2015spin} and is studied in a system formed by a non-spherical satellite of negligible mass (modeled as a dumbbell) orbiting around a non-spherical mass distribution (modeled as an ellipsoid). The analysis on the asteroidal systems (87) Sylvia and (216) Kleopatra show spin-spin resonances but there is no evidence of chaos in their motions. 

The discovery of Pluto's small moons in the last decade has resulted in much attention directed to understanding the dynamics of Pluto's satellites. The four small moons (Styx, Nix, Kerberos, and Hydra) orbit around the binary system formed by Pluto and its largest moon, Charon. \cite{showalter2015resonant} looked for resonances in the orbits of Pluto's small moons and they found that the Pluto-Charon system appears to induce chaos in non-spherical moons without the need of resonance, which was not seen in other well studied three-body systems such as Saturn-Titan-Hyperion. In particular, \cite{showalter2015resonant} conjectured that non-spherical moons may rotate chaotically with no resonance required. An orbit classification on the Pluto-Charon system is provided in \cite{zotos2015orbit}, in which the author, using numerical simulations on the gravitational potential generated by the binary system, distinguishes three types of orbits (bounded, escaping, and collisional) depending on their initial conditions. In \cite{correia2015spin}, the spin-orbit coupling for a non-spherical satellite orbiting around a binary system was studied, with a particular application to the small moons of the Pluto-Charon system. Their numerical simulations show chaotic behavior for the small moons of Pluto. In particular, a dynamical mechanism for the chaotic behavior was proposed by \cite{correia2015spin} who showed that for slowly spinning satellites, spin-binary resonances from Charon's periodic perturbations are sufficiently strong to cause chaotic tumbling. Finally, in \cite{jiang2016dynamical}, the authors study the $N$-body problem for irregular bodies, including gravitational, electric, and magnetic fields. They present numerical simulations on several $N$-body systems, with the full six-body Pluto's system among them.

Recent New Horizons observations showed that the low mass satellites are spinning faster than considered by \cite{showalter2015resonant,correia2015spin}, with angular spin rates many times greater than their orbital mean motions, implying that despinning due to tidal dissipation has not taken place \cite{weaver2016small}. At higher spin states, spin-orbit and spin-binary resonances may not be as strong, so chaotically tumbling is not assured. In \cite{quillen2017obliquity}, the authors used a damped mass-spring model within an $N$-body simulation to study spin and obliquity evolution for single spinning non-round bodies in circumbinary orbit. Their simulations with tidal dissipation alone do not show strong obliquity variations from tidally induced spin-orbit resonance crossing and this was attributed to the high satellite spin rates and low orbital eccentricities. However, a tidally evolving Styx does exhibit intermittent obliquity variations and episodes of tumbling. As stated in \cite{quillen2017obliquity}, ``\textit{We lack simple dynamical models for phenomena seen in our simulations, such as excitation of tumbling.}" 

With such applications in mind, we will consider an $N$-body system in which all of the bodies are point masses, except for a single body which has finite size. We then present a reduced model which considers the body as a 1D geometry. The body is treated as a rod with an arbitrary mass distribution in this simplified model. This model contrasts with other work which consider the gravitational field as a perturbation from a perfect sphere (i.e. ellipticity close to zero), such as the work presented in \cite{Correia2014,Delisle2017}. Our 1D rod model is highly asymmetric, equivalent to an ellipticity of unity, and is an approximation for celestial bodies such as the interstellar asteroid `Oumuamua which was recently observed \cite{meech2017brief}. Additionally, our rotational equation of motion for the 1D rod is valid for non-Keplerian orbits which is in contrast with the Beletsky equation \cite{Beletskii1966,CellettiBook2010} for the rotation of dumbbell systems, of which our model can be considered as a continuum limit.

The translational motion equations in our model are equivalent to the $N$-body problem for point masses whilst an additional rotational motion equation, arising from tidal corrections to the gravitational force, is included for the rigid body. Thus, if orbits are known, they can be fed into the rotational motion equation which can then be solved. The rotational motion equation only depends on the positions of the other point masses. Note that since the translational motion equations do not depend on the angles or inertial terms of the rigid body, chaos in the rotational motion of one or more bodies will not result in chaotic trajectories of the orbits themselves. Such dynamics give a possible explanation for the route to chaotic dynamics observed in $N$-body systems such as the Pluto system. Our results here emphasize that tidal dissipation, obliquity variation, and orbital resonances are not required for the existence of chaotic rotations in the simple planar setting. These results suggest generic chaotic rotation of non-spherical moons in more realistic settings, as conjectured by \cite{showalter2015resonant}. 

The rest of the paper is organized as follows: In Section 2, we shall describe the simplified model of the $N$-body problem including rotation of a single rigid body which has a 1D geometry. In Section 3, we apply the Melnikov method in order to analytically demonstrate the emergence of homoclinic chaos in the rotational motion equation under perturbations of approximately circular orbits. This analytical approach requires the construction of homoclinic connections which is possible in fairly restricted cases, such as that of the circular orbit. Therefore to consider more complicated orbits, we provide numerical simulations and consider resulting time series and Poincar\'e sections in Section 4 to demonstrate the existence of chaotic dynamics in the rotational motion of the rigid bodies under perturbations of more complicated orbits, such as the well-known figure-8 orbit. Such results suggest that chaos from perturbations of orbits in the $N$-body problem ($N \geq 2$) for a rigid body is ubiquitous. We summarize and discuss the results in Section 5.

\section{Planar orbital dynamical system of a single rotating rod}

We study a system of $N$ bodies that exert gravitational forces on each other, with $N-1$ point masses and a single rigid body of finite geometry, which we will refer to as the $N$th body hereafter. In particular, we consider a rigid body in the shape of a 1D rod. We view this as a proxy for a more general asymmetric body whose geometry deviates largely from the sphere, but allows for a number of analytical simplifications, permitting later analysis of the dynamics. The orbital dynamics are taken to be planar as this both simplifies the mathematics yet still agrees with many real orbital systems, such as the Pluto system. Additionally, instability in the planar setting suggests instability in the full problem as there are more degrees of freedom.

We suppose the motion, both translational and rotational, of the bodies is confined to the $x$-$y$ inertial plane and we do not assume that the orbit is necessarily Keplerian. We denote the position of the $i$th body's center of mass by the vector $\boldsymbol{r}_{i}=\left(x_{i}, y_{i}\right)$ and the orientation of the $N$th rigid rod with respect to the positive $x$-axis as the angle $\theta$. The equations of orbital motion and additional rotation due to the gravitational field generated by the $N$ bodies, which are separated on a length-scale that is much larger than the size of the rigid rod, is given by (see Appendix A for details)
\begin{align}
\frac{\mathrm{d}^{2}x_{i}}{\mathrm{d}t^{2}} & =\sum_{j\neq i}\frac{Gm_{j}\left(x_{j}-x_{i}\right)}{\left\{ \left(x_{j}-x_{i}\right)^{2}+\left(y_{j}-y_{i}\right)^{2}\right\} ^{3/2}},\label{eq:Dynamicalsystemstart}\\
\frac{\mathrm{d}^{2}y_{i}}{\mathrm{d}t^{2}} & =\sum_{j\neq i}\frac{Gm_{j}\left(y_{j}-y_{i}\right)}{\left\{ \left(x_{j}-x_{i}\right)^{2}+\left(y_{j}-y_{i}\right)^{2}\right\} ^{3/2}},
\end{align}
for $i=1,2,\dots, N$, and
\begin{equation}
\frac{\mathrm{d}^2 \theta}{\mathrm{d}t^2} = \frac{3G}{2}\sum_{j=1}^{N-1}\frac{m_j\left(2\left(x_{j}-x_{N}\right)\left(y_{j}-y_{N}\right)\cos2\theta+\left[\left(y_{j}-y_{N}\right)^{2}-\left(x_{j}-x_{N}\right)^{2}\right]\sin2\theta\right)}{\left(\left(x_{j}-x_{N}\right)^{2}+\left(y_{j}-y_{N}\right)^{2}\right)^{5/2}}.
\label{eq:Dynamicalsystemend}
\end{equation}

Scaling the angular variable as $\hat{\theta} = 2\theta$ and then dropping hats, we have that the rotational variable $\theta$ evolves via a non-autonomous ODE of the form
\begin{equation}
\frac{\mathrm{d}^{2}\theta}{\mathrm{d}t^{2}}  = F_{1}(t)\cos\theta - F_{2}(t)\sin\theta,
\end{equation}
where
\begin{equation}\label{fy}
F_{1}(t) = \sum_{j=1}^{N-1}\frac{6Gm_{j} \left(x_{j}-x_{N}\right)\left(y_{j}-y_{N}\right)}{\left\{ \left(x_{j}-x_{N}\right)^{2}+\left(y_{j}-y_{N}\right)^{2}\right\} ^{5/2}},
\end{equation}
and 
\begin{equation}\label{fx}
F_{2}(t) = \sum_{j=1}^{N-1}\frac{3Gm_{j} \left[\left(y_{j}-y_{N}\right)^{2}-\left(x_{j}-x_{N}\right)^{2}\right]}{\left\{ \left(x_{j}-x_{N}\right)^{2}+\left(y_{j}-y_{N}\right)^{2}\right\} ^{5/2}}.
\end{equation}

We comment on some important aspects of \eqref{eq:Dynamicalsystemstart}-\eqref{eq:Dynamicalsystemend}. First,  the translational dynamics of all masses in the system are given by the point approximation which is a consequence of the rigid body having a highly asymmetric geometry that is very small compared to the separations of all the bodies. In particular, the quadrupole or first order corrections to the gravitational field generated and felt by the $N$th rigid body vanish. This aspect contrasts with other models which consider the bodies as perturbations from perfect spheres, such as the work presented in \cite{Correia2014} where such corrections are non-trivial in the orbital dynamics. Other instances where the higher order terms are non-negligible include the opposite regime considered here where the bodies are no longer well separated with respect to their size \cite{Delisle2017}.  

Second, the rotational equation of motion \eqref{eq:Dynamicalsystemend} is valid for non-Keplerian orbits which is in contrast with the Beletsky equation \cite{Beletskii1966,CellettiBook2010} for the rotation of dumbbell systems. Note that this dumbbell system is a singular limit of our model presented here. However, in our case the angular equation completely decouples from the translational equations, meaning that the $2N$ translational equations can be solved and their solutions placed into the rotation equation. 

\section{Application of the Melnikov method for homoclinic chaos}
We consider the Melnikov method \cite{mel1963stability,arnold1964instability,guckenheimer2013nonlinear,wiggins2003introduction} in order to show that system \eqref{eq:Dynamicalsystemstart}-\eqref{eq:Dynamicalsystemend} admits homoclinic chaos, as the method is used to understand chaos resulting from nonintegrable perturbations of Hamiltonian systems \cite{holmes1982melnikov,holmes1982horseshoes,camassa1998melnikov,yagasaki1999method}. We note that \eqref{eq:Dynamicalsystemstart}-\eqref{eq:Dynamicalsystemend} is of this near-integrable form, with the dominant integrable part comprising of the orbital dynamics of the point masses and the constant rotation of the rigid body, given that the tidal forces vanish to leading order. The nonintegrable part of the Hamiltonian comes from the gravitational torque applied to the body, specifically for orbits which are non-circular as will be shown, similar to the Beletsky equation for non-zero orbit eccentricity \cite{Burov1984,CellettiSidorenko2008}.  The approach has been used to find homoclinic or heteroclinic chaos in a variety of systems \cite{lin1990using,hogan1992heteroclinic,yagasaki1994homoclinic,spyrou2000nonlinear,zhang2009using,zhang2010extended} and has also been used to assist in the control of such chaos \cite{chacon2006melnikov}. The method has been applied to homoclinic orbits corresponding to elliptic functions \cite{belhaq2000homoclinic}. The approach has also been used in the study of nonlinear PDEs, such as a reduction of the perturbed KdV equation \cite{grimshaw1994periodic}. Of immediate relevance to our problem, in addition to the recent application to the dumbbell satellite problem \cite{fernandez2016dynamics}, the method has been applied to restricted three-body problems \cite{xia1992melnikov}.

We show that chaos occurs in a system consisting of a circular orbit perturbed under small deformations. To do so, we apply Melnikov's method, which can be used to prove the existence of transverse homoclinic orbits in the non-autonomous dynamical system
\begin{equation}
\dot{\boldsymbol{u}}=\boldsymbol{f}\left(\boldsymbol{u}\right)+\eta\boldsymbol{g}\left(\boldsymbol{u},t\right),\label{eq:GeneralMelnikov}
\end{equation}
for $\eta$ sufficiently small. In our setting, \eqref{eq:GeneralMelnikov} is the first-order form of the second order equation \eqref{eq:Dynamicalsystemend}. Note that this method is valid for non-automonous perturbations which are periodic \cite{mel1963stability}, quasiperiodic \cite{Wiggins1987}, and random \cite{LuWang2010} and there is no assumption made as to the frequency of such perturbations. 

The quantity of interest is the Melnikov function, which takes the
form
\begin{equation}
M\left(t_{0}\right)=\intop_{-\infty}^{\infty}\boldsymbol{f}\left(\boldsymbol{u}^{\left(0\right)}\left(t\right)\right)\wedge\boldsymbol{g}\left(\boldsymbol{u}^{\left(0\right)}\left(t\right),t+t_{0}\right)\mathrm{d}t,\label{eq:Melnikovfunction}
\end{equation}
where $\boldsymbol{u}^{\left(0\right)}\left(t\right)$ is the explicit form of a homoclinic orbit for $\eta=0$ in (\ref{eq:GeneralMelnikov}), $t_{0}\in\mathbb{R}$, and $\boldsymbol{f}\wedge\boldsymbol{g} = f_1 g_2-f_2 g_1$. The Melnikov function quantifies the distance between unstable and stable manifolds, so if there exists a $t_{0}$ such that there is a simple zero of $M\left(t_{0}\right)$, then transverse homoclinic
orbits exist in the dynamical system. Furthermore, such trajectories
imply the existence of a Smale horseshoe and, by extension, chaos \cite{guckenheimer2013nonlinear}.

For the present analysis, we study the planar rotation of a single
body, which, in its simplest form, is described by the non-automonous
system
\begin{equation}\label{thetaeqnsimplify}
\frac{\mathrm{d}^{2}\theta}{\mathrm{d}t^{2}}=F_{1}\left(t\right)\cos\theta-F_{2}\left(t\right)\sin\theta,
\end{equation}
where $F_{1}\left(t\right)$ and $F_{2}\left(t\right)$ (as given in \eqref{fy}-\eqref{fx}) are forces acting on the rod and will generally
depend on time.

\subsection{Purely circular orbit and forcing}
Consider the case where the orbit of the $N$th rigid body is circular, while the other orbits are small relative to this circular orbit. Mathematically, we write $x_N = -A\cos(\gamma t/2)$, $y_N = -A\sin(\gamma t/2)$, with $x_j = A\delta P_j(t)$ and $y_j = A\delta Q_j(t)$ for periodic $P_j(t),Q_j(t)$ and $\delta << 1$ for all $j = 1,2,\dots, N-1$. Here $\gamma>0$ is a parameter related to the period of the orbit. Taking these in \eqref{eq:Dynamicalsystemend} and applying multiple-angle trigonometric identities, we then have
\begin{equation}\begin{aligned}
\frac{\mathrm{d}^{2}\theta}{\mathrm{d}t^{2}}   & = \sigma (\sin(\gamma t)\cos(\theta)-\cos(\gamma t)\sin(\theta) )  \\
&  \qquad +  \frac{3\sigma \delta}{2}  \sin(\theta)\sum_{j=1}^{N-1} m_{j} \left\lbrace (5\cos(3\gamma t/2)+\cos(\gamma t/2))P_j(t) \right.\\
& \qquad\qquad \qquad\qquad \qquad \left. + (5\sin(3\gamma t/2) -\sin(\gamma t/2))  Q_j(t) \right\rbrace\\
&   \qquad +  \frac{3\sigma \delta}{2}  \cos(\theta)\sum_{j=1}^{N-1} m_{j} \left\lbrace  (5\cos(3\gamma t/2)-\cos(\gamma t/2))Q_j(t)\right.\\
& \qquad\qquad \qquad\qquad\qquad  \left. - (5\sin(3\gamma t/2) + \sin(\gamma t/2))  P_j(t)  \right\rbrace\\
& \qquad  + O(\delta^2)\,,
\end{aligned}\end{equation}
where 
\begin{equation}
\sigma = \frac{3(N-1)G}{A^3}\,.
\end{equation}

From the above derivation, if the orbit of the rotating rod is purely circular, then the forces in the $x$ and $y$ direction will be sinusoidally periodic with
a non-zero angular frequency $\gamma$ such that their explicit form is
\begin{equation}
\boldsymbol{F} = 
\left(\begin{matrix}F_{x}\\
F_{y}\end{matrix}\right) 
= \sigma \left(\begin{matrix}\cos(\gamma t)\\
\sin(\gamma t)\end{matrix}\right) ,
\end{equation}
up to lowest order (neglecting $O(\delta)$ corrections). The rotation of the rod then evolves according to the non-autonomous ODE
\begin{equation}
\frac{\mathrm{d}^{2}\theta}{\mathrm{d}t^{2}}=\sin(\gamma t)\cos(\theta)-\cos(\gamma t)\sin(\theta) =-\sin\left(\theta-\gamma t\right),\label{eq:Circular motion}
\end{equation}
where the time and angular frequency have been rescaled as $t\rightarrow t\sqrt{\sigma}$ and $\gamma\sqrt{\sigma}\rightarrow \gamma$. Making the change of variable $\tilde{\theta}=\theta-\gamma t$, which relates the old and new angular variable by a constant rotation of angular velocity $\gamma$, we find that (\ref{eq:Circular motion})
reduces to 
\begin{equation}
\frac{\mathrm{d}^{2}\tilde{\theta}}{\mathrm{d}t^{2}}=-\sin\tilde{\theta},
\label{eq:Circularrescaled}
\end{equation}
which is an automonous system in $\tilde{\theta}$. This result implies
that no aperiodicity will occur in $\theta$ and, as such, a rod cannot
exhibit chaotic rotation if it is in a purely circular orbit.

We note that (\ref{eq:Circularrescaled}) has exactly one homoclinic
orbit, given the periodic domain in $\tilde{\theta}\in\left(-\pi,\pi\right]$.
To determine its form, we multiply (\ref{eq:Circularrescaled}) through
by $\mathrm{d}\tilde{\theta}/\mathrm{d}t$ and integrate with respect
to $t$ to obtain 
\begin{equation}
\frac{1}{2}\left(\frac{\mathrm{d}\tilde{\theta}}{\mathrm{d}t}\right)^{2}+V\left(\tilde{\theta}\right)=E,
\end{equation}
where $E$ is a constant analogous to the total mechanical energy
of the rotating rod in a potential $V\left(\tilde{\theta}\right)=-\cos\tilde{\theta}$.
The homoclinic orbit connects the fixed point at $\tilde{\theta}=\pi$
along the level set $E=1$, so the homoclinic orbit in $\tilde{\theta}^{\left(0\right)}$
is found by solving 
\begin{equation}
\frac{\mathrm{d}\tilde{\theta}^{\left(0\right)}}{\mathrm{d}t}=\pm\sqrt{2\left(1+\cos\tilde{\theta}^{\left(0\right)}\right)},
\end{equation}
for the initial condition $\tilde{\theta}^{\left(0\right)}\left(t=0\right)=0$. Using a double angle formula, 
$\frac{\mathrm{d}\tilde{\theta}^{\left(0\right)}}{\mathrm{d}t}=\pm2\cos\left(\tilde{\theta}/2\right)$,
and, by introducing a rescaling $\tilde{\theta}^{\left(0\right)}/2\rightarrow\phi^{\left(0\right)}$,
we obtain, by separation of variables, 
$\ln\left|\sec\phi^{\left(0\right)}+\tan\phi^{\left(0\right)}\right|  =\pm t$,
having imposed the initial condition $\phi^{\left(0\right)}\left(t=0\right)=0$. Rearranging to find an explicit form for $\phi^{\left(0\right)}$, $\sec\phi^{\left(0\right)}+\tan\phi^{\left(0\right)}=\exp\left(\pm t\right)$. We simplify this expression to $\tan\phi^{\left(0\right)}  =\frac{1}{2}\left\{ \exp\left(\pm t\right)-\exp\left(\mp t\right)\right\}$, 
which results in the time explicit angular position
\begin{equation}
\tilde{\theta}^{\left(0\right)}\left(t\right)=\pm2\tan^{-1}\left(\sinh\left(t\right)\right).\label{eq:Homoclinictheta}
\end{equation}
Note that $\tan^{-1}\left(t\right)$ is an odd function, and that (\ref{eq:Homoclinictheta}) approaches $\pm\pi$ as $t\rightarrow\pm\infty$ for the positive solution, and the opposite sign for the negative solution. By differentiating with respect to time, we find the time explicit
angular velocity 
\begin{equation}
\frac{\mathrm{d}}{\mathrm{d}t}\tilde{\theta}^{\left(0\right)}=\pm2\mathrm{sech}\left(t\right).\label{eq:Homoclinicthetadot}
\end{equation}

\subsection{Perturbations of the circular forcing}

We now consider an arbitrary number of periodic perturbations to the circular orbit which are of $O\left(\eta\right)$ such that $\delta \ll \eta \ll 1$.
Explicitly, suppose that the force exerted on the rod takes the form
\begin{equation}
\boldsymbol{F} = 
\sigma \left(\begin{matrix}\cos(\gamma t)\\
\sin(\gamma t)\end{matrix}\right)
+ \sigma \eta \left(\begin{matrix}\mathcal{F}_{11}(t)\\
\mathcal{F}_{2}(t)\end{matrix}\right)\,,
\end{equation}
where $\mathcal{F}_{1},\mathcal{F}_{2}$ are bounded and $C^1(\mathbb{R})$ perturbations. We shall assume that such perturbations are oscillatory about zero (including periodic or quasi-periodic perturbations) and in practice write
\begin{equation}
\left(\begin{matrix}\mathcal{F}_{1}(t)\\
\mathcal{F}_{2}(t)\end{matrix}\right) =  \sum_{k=1}^{\infty} \left(\begin{matrix}  \alpha_k\sin (kpt) + \beta_k \cos (kpt)  \\
 \hat{\alpha}_k\sin (kpt) + \hat{\beta}_k \cos (kpt) \end{matrix}\right),
\end{equation}
where the coefficients $\alpha_{k}$, $\beta_{k}$, $\hat{\alpha_{k}}$, and $\hat{\beta_{k}}$ are $O(1)$ in and $p>0$ is a constant. The perturbation angular frequencies $p\in\mathbb{R}$ need not necessarily be integer multiples of $\gamma$ or even rational multiples, allowing us to consider not only periodic but also quasi-periodic orbits with precession, provided $\eta\ll1$ holds. In the case of periodic forcing, the series will terminate at some maximal index $k$, resulting in finite sums. In the quasi-periodic forcing case, infinitely many terms would be needed, and one should ensure that the perturbation series do converge in such cases. For our interests, we shall consider only perturbations with convergent Fourier series.

The non-autonomous evolution of the angular position then reads 
\begin{equation}\begin{aligned}
\frac{\mathrm{d}^{2}\theta}{\mathrm{d}t^{2}} & =-\sin\left(\theta-\gamma t\right)+\eta\cos\theta \sum_{k=1}^{\infty} \left\lbrace \hat{\alpha}_k\sin (kpt) + \hat{\beta}_k \cos (kpt) \right\rbrace \\
& \qquad\qquad -\eta\sin\theta \sum_{k=1}^{\infty} \left\lbrace\alpha_k\sin (kpt) + \beta_k \cos (kpt) \right\rbrace ,
\end{aligned}\end{equation}
where we have rescaled time as $t\rightarrow t\sqrt{\sigma}$ and $p\sqrt{\sigma}\rightarrow p$. Making the change of variables $\tilde{\theta}=\theta-\gamma t$, applying trigonometric rules, and assuming the necessary results on the convergence of the series, we obtain
\begin{equation}\begin{aligned}
\frac{\mathrm{d}^{2}\tilde{\theta}}{\mathrm{d}t^{2}} & =-\sin\left(\tilde{\theta}\right)  \\
& \qquad +\frac{\eta}{2}\sum_{k=1}^{\infty} \left\lbrace (\hat{\alpha}_{k} - \beta_k) \sin\left(\tilde{\theta}+\left(\gamma+kp\right)t\right)
- (\hat{\alpha}_k + \beta_k) \sin\left(\tilde{\theta}+\left(\gamma-kp\right)t\right) \right.\\
& \qquad\qquad \left. + (\hat{\beta}_k + \alpha_k ) \cos\left(\tilde{\theta}+\left(\gamma+kp\right)t\right)
+ (\hat{\beta}_k - \alpha_k) \cos\left(\tilde{\theta}+\left(\gamma-kp\right)t\right) \right\rbrace .\label{eq:Generalperturbedequation}
\end{aligned}\end{equation}

In the particular case of (\ref{eq:Generalperturbedequation}), equation (\ref{eq:Melnikovfunction})
becomes
\begin{equation}\begin{aligned}
M\left(t_{0}\right) & =\frac{1}{2}\intop_{-\infty}^{\infty} \frac{\mathrm{d}}{\mathrm{d}t}\left(\tilde{\theta}^{\left(0\right)}\left(t\right)\right)   \sum_{k=1}^{\infty}\left\lbrace (\hat{\beta}_k + \alpha_k) \cos\left(\tilde{\theta}^{\left(0\right)}+\left(\gamma+kp\right) \left(t+t_{0}\right)\right) \right.\\
 & \qquad\qquad  + (\hat{\beta}_k - \alpha_k) \cos\left(\tilde{\theta}^{\left(0\right)}+
 \left(\gamma-kp\right)\left(t+t_{0}\right)\right)   \\
 & \qquad\qquad + (\hat{\alpha}_k  - \beta_k)\sin\left(\tilde{\theta}^{\left(0\right)}+\left(\gamma+kp\right)\left(t+t_{0}\right)\right)\\
 & \qquad\qquad \left. - (\hat{\alpha}_k + \beta_k)\sin\left(\tilde{\theta}^{\left(0\right)}+\left(\gamma-kp\right)\left(t+t_{0}\right)\right)
 \right\rbrace \mathrm{d}t,\label{eq:ParticularMelnikovfunction}
\end{aligned}\end{equation}
where $\tilde{\theta}^{\left(0\right)}\left(t\right)$ and $\frac{\mathrm{d}\tilde{\theta}^{\left(0\right)}}{\mathrm{d}t}$
are the explicit homoclinic orbits of the rescaled angular position
and velocity in the purely circular system (\ref{eq:Circularrescaled}).

Using (\ref{eq:Homoclinictheta}) and (\ref{eq:Homoclinicthetadot}) in the particular Melnikov function (\ref{eq:ParticularMelnikovfunction}) and performing the integration, we obtain 
\begin{equation}\begin{aligned}
M\left(t_{0}\right) & =\pm \sum_{k=1}^{\infty} \left\lbrace \mu_{k,+} (\hat{\beta}_k + \alpha_k)  \cos\left[\left(\gamma+kp\right)t_{0}\right] + \mu_{k,-} (\hat{\beta}_k - \alpha_k)  \cos\left[\left(\gamma-kp\right)t_{0}\right] \right.\\
& \qquad\qquad \left. + \mu_{k,+} (\hat{\alpha}_k  - \beta_k) \sin\left[\left(\gamma+kp\right)t_{0}\right] - \mu_{k,-}(\hat{\alpha}_k + \beta_k) \sin\left[\left(\gamma-kq\right)t_{0}\right] \right\rbrace ,\label{eq:ExactMelnikovfunction}
\end{aligned}\end{equation}
where
\begin{equation}
\mu_{k,\pm} = \pi \left(\gamma \pm kp\right)^{2}\left\{ \mathrm{sech}\left[\frac{\pi}{2}\left(\gamma \pm kp\right)\right] - \mathrm{csch}\left[\frac{\pi}{2}\left(\gamma \pm kp\right)\right]\right\}.
\end{equation}

Recall that we assume $\mathcal{F}_{1},\mathcal{F}_{2}$ are bounded and $C^1(\mathbb{R})$ perturbations which are oscillatory about zero. Define 
\begin{equation}\label{muhat}
\hat{\mu}(\chi) = \pi \chi^2 \left\{ \mathrm{sech}\left[\frac{\pi\chi}{2}\right] - \mathrm{csch}\left[\frac{\pi\chi}{2}\right]\right\}\,,
\end{equation}
and note that $\hat{\mu}(0) = 0$, $\exp(|\chi|)\hat{\mu}(\chi) \rightarrow 0$ as $\chi \rightarrow \pm \infty$, and $|\hat{\mu}(\chi)|< \overline{\mu} \approx 2.76$ for all $\chi \in \mathbb{R}$, where this value comes from extremizing \eqref{muhat}. Therefore, $\mu_{k,\pm} = 0$ if and only if $\gamma \pm kp =0$, and the sequences $\mu_{k,\pm}$ are bounded and decay for large enough $k$ (with the decay rate being faster than exponential).

Manipulating the series in \eqref{eq:ExactMelnikovfunction}, we find that 
\begin{equation}\begin{aligned}
M\left(t_{0}\right) & = - \sin(\gamma t_0) \sum_{k=1}^\infty \left( \mu_{k,+}+\mu_{k,-} \right)\left( \alpha_k \sin(kpt_0)+ \beta_k \cos(kpt_0)\right) \\
& \qquad\qquad + \cos(\gamma t_0) \sum_{k=1}^\infty \left( \frac{\mu_{k,+}-\mu_{k,-}}{kp} \right)\left( \alpha_k \sin(kpt_0)+ \beta_k \cos(kpt_0)\right)'\\
& \qquad\qquad + \cos(\gamma t_0) \sum_{k=1}^\infty \left( \mu_{k,+}+\mu_{k,-} \right)\left( \hat{\alpha}_k \sin(kpt_0)+ \hat{\beta}_k \cos(kpt_0)\right)\\
& \qquad\qquad - \sin(\gamma t_0) \sum_{k=1}^\infty \left( \frac{\mu_{k,+}-\mu_{k,-}}{kp} \right)\left( \hat{\alpha}_k \sin(kpt_0)+ \hat{\beta}_k \cos(kpt_0)\right)'\,,
\end{aligned}\end{equation}
and recalling the definitions of $\mathcal{F}_{1},\mathcal{F}_{2}$, we see that $M(t_0)$ involves terms which are scalings of the terms in $\mathcal{F}_{1},\mathcal{F}_{2}$ and their derivatives. Due to the properties of $\hat{\mu}(\chi)$ listed above, we have that
\begin{equation}\begin{aligned}
\left| M(t_0) \right| & \leq 2\left(\max_{\chi}\hat{\mu}(\chi) \right)\max_{t\in \mathbb{R}}\left\lbrace \left| \mathcal{F}_{1}(t)\right| + \frac{1}{p}\left| \mathcal{F}_{1}'(t)\right| + \left| \mathcal{F}_{2}(t)\right| + \frac{1}{p}\left| \mathcal{F}_{2}'(t)\right|\right\rbrace \\
& \leq 2\overline{\mu}\max_{t\in \mathbb{R}}\left\lbrace \left| \mathcal{F}_{1}(t)\right| + \frac{1}{p}\left| \mathcal{F}_{1}'(t)\right| + \left| \mathcal{F}_{2}(t)\right| + \frac{1}{p}\left| \mathcal{F}_{2}'(t)\right|\right\rbrace \,.
\end{aligned}\end{equation}
As $p$ is positive and finite, and $\mathcal{F}_{1},\mathcal{F}_{2}$ are bounded with bounded derivatives, then the function $M(t_0)$ is bounded in $t_0$. Since $\mathcal{F}_{1},\mathcal{F}_{2}$ are oscillatory about zero, so are their derivatives. Then either $M(t_0)$ is oscillatory about zero or it is a constant. The only possible constant value is zero, and if this occurs then $M(t_0)$ is identically zero and we have no simple zeros.

Note that for $\gamma , p >0$, then at most one of $\mu_{k,+}, \mu_{k,-}$ can be zero. If there is more than one term in the perturbation expansion, then there will always be at least one non-zero coefficient of a sine or cosine term, and hence $M(t_0)$ will have countably many simple zeros. If there is exactly one term in each perturbation expansion, then $M(t_0)$ will have countably many simple zeroes unless $\gamma = -p$, $\hat{\beta}_1 = \alpha_1$, $\hat{\alpha}_1=-\beta_1$ or $\gamma = p$, $\hat{\beta}_1 = -\alpha_1$, $\hat{\alpha}_1=\beta_1$ (where, without loss of generality, $k=1$ in a one-term perturbation expansion). 

Therefore, excluding such a narrow class of perturbations, the function $M(t_0)$ is not identically zero and oscillates about zero. As such, there exist countably infinitely many values of $t_{0}\in\mathbb{R}$ for which simple zeros will occur regardless of the values for $\gamma$, again provided that the coefficients of \eqref{eq:ExactMelnikovfunction} are not such that $M(t_0) \equiv 0$. By the Melnikov criteria, transverse homoclinic orbits exist in the perturbed periodic system for $\eta\ll1$ under such conditions.

These results suggest that the physical parameters associated with the orbital dynamics and the rotational dynamics are irrelevant. The only necessary requirement for chaos is that the gravitational torque exerted on the rigid body is non-zero. Note however that this result only shows the existence of chaotic rotation in the system, and not the extent to which it occurs. For example, rotational periodic motions may still be possible \cite{levin1993dynamics,CellettiSidorenko2008}.

\section{Numerical simulations and verification of chaos}
While the results obtained from the Melnikov method demonstrate that chaos should be ubiquitous due to arbitrary small perturbations of circular orbits, for more complicated orbits the analytical method is not tractable. As such, we shall consider numerical simulations in order to demonstrate chaotic dynamics in a variety of other orbital configurations, which suggests that chaotic rotations should be quite common in planar $N$-body dynamics. We shall begin by showing that the numerical simulations agree with the Melnikov analysis in the case of a single nearly circular orbit, in order to verify the numerical approach against our analytical results. Our focus then shifts to three bodies in more complicated orbital configurations, which suggests that the results are robust for $N$-body dynamics.

\begin{figure}
\includegraphics[width=0.49\textwidth]{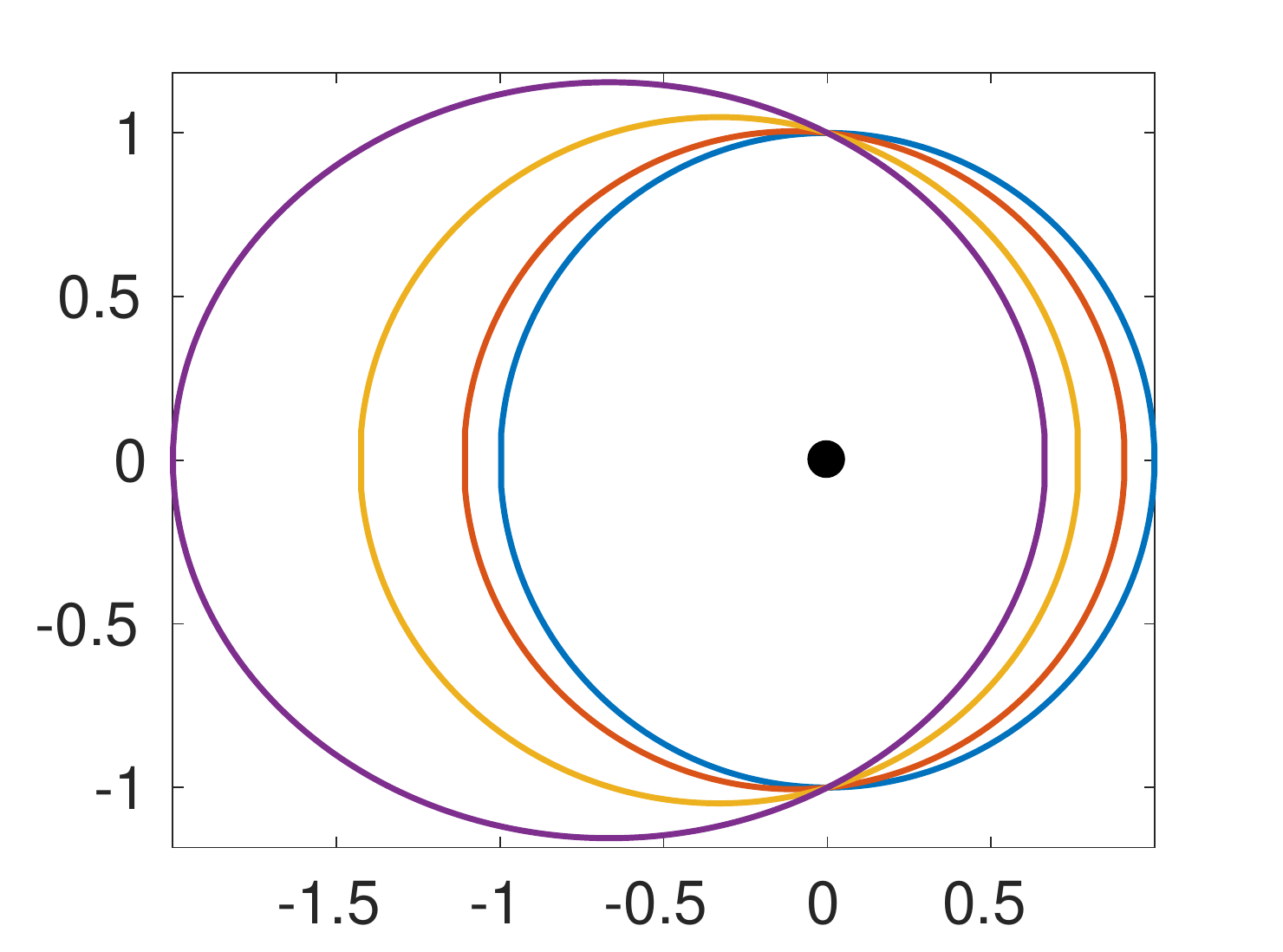}
\includegraphics[width=0.49\textwidth]{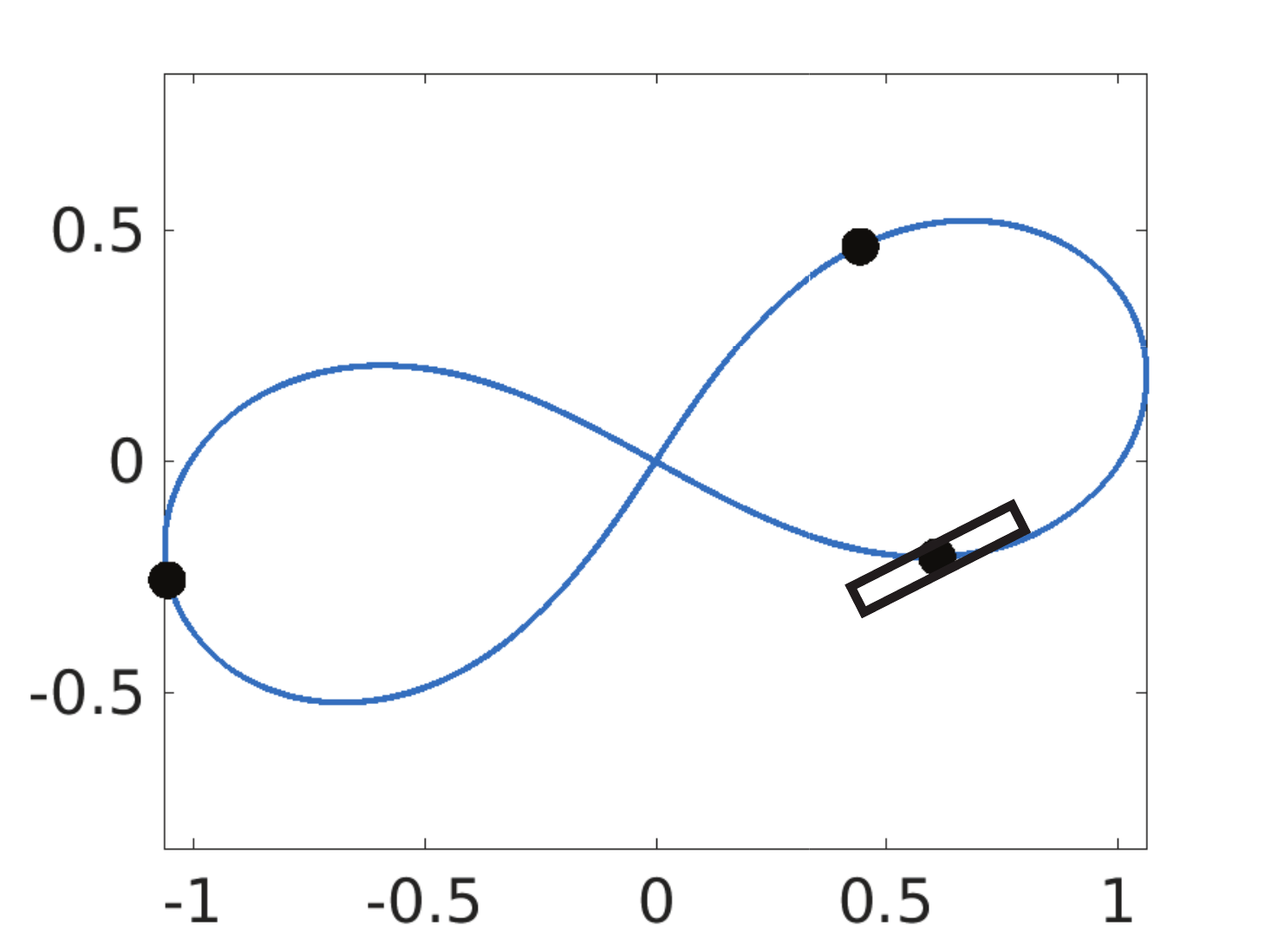}
\caption{The plot on the left depicts four different circular orbits for the rigid rod around a massive body with initial perturbations of $\dot{y}_1(0)=0,0.1,0.3$, and $0.5$ leading to successively larger circular orbits that are to the left of the massive body (which remains approximately fixed independent of the perturbation). The plot on the right is of a figure-8 orbit with three bodies of identical masses.}\label{Orbits}
\end{figure}

We now demonstrate numerical simulations of rotating rigid bodies within sample orbital simulations, and discuss the transition to chaotic dynamics observed. We study two examples in detail: perturbed circular orbits in a two-body system, and the figure-8 orbit of the three body system. Both of these orbits are shown in Fig. \ref{Orbits}. The simulations were implemented using the \textsc{Matlab} function \textsc{ode113} with relative and absolute tolerances set to $2.22045\times 10^{-14}$. We also performed the same simulations using an 8-stage Runge-Kutta scheme due to \cite{Papakostas} in \textsc{Matlab}, as well as in \textsc{Mathematica} using the function \textsc{NDSolve} and found consistent results using all three methods for all simulations shown.

\subsection{Simulations for perturbed circular orbits}
To generate our circular orbits in the two-body case we set $G=1$, $m_1=1$, and $m_2=10^{-13}$ so that the second body will have a negligible effect on the first, which will be stationary to within an approximately-circular orbit of size $O(10^{-13})$. We set $x_2(0)=0$, $y_2(0)=1$, $x_1(0)=0$, $y_1(0)=0$, $\dot{x}_2(0) = -(m_1^2/(m_1+m_2))^{1/2} \approx -1$, $\dot{x}_1(0) = - \dot{x}_2(0)$, $\dot{y}_1(0) = 0$, and vary $\dot{y}_2(0)$ as a perturbation or bifurcation parameter. We simulate the system for $T=300$ units of time corresponding to $\approx 550$ orbits (which depends on the perturbed orbit).

\begin{figure}
\includegraphics[width=0.49\textwidth]{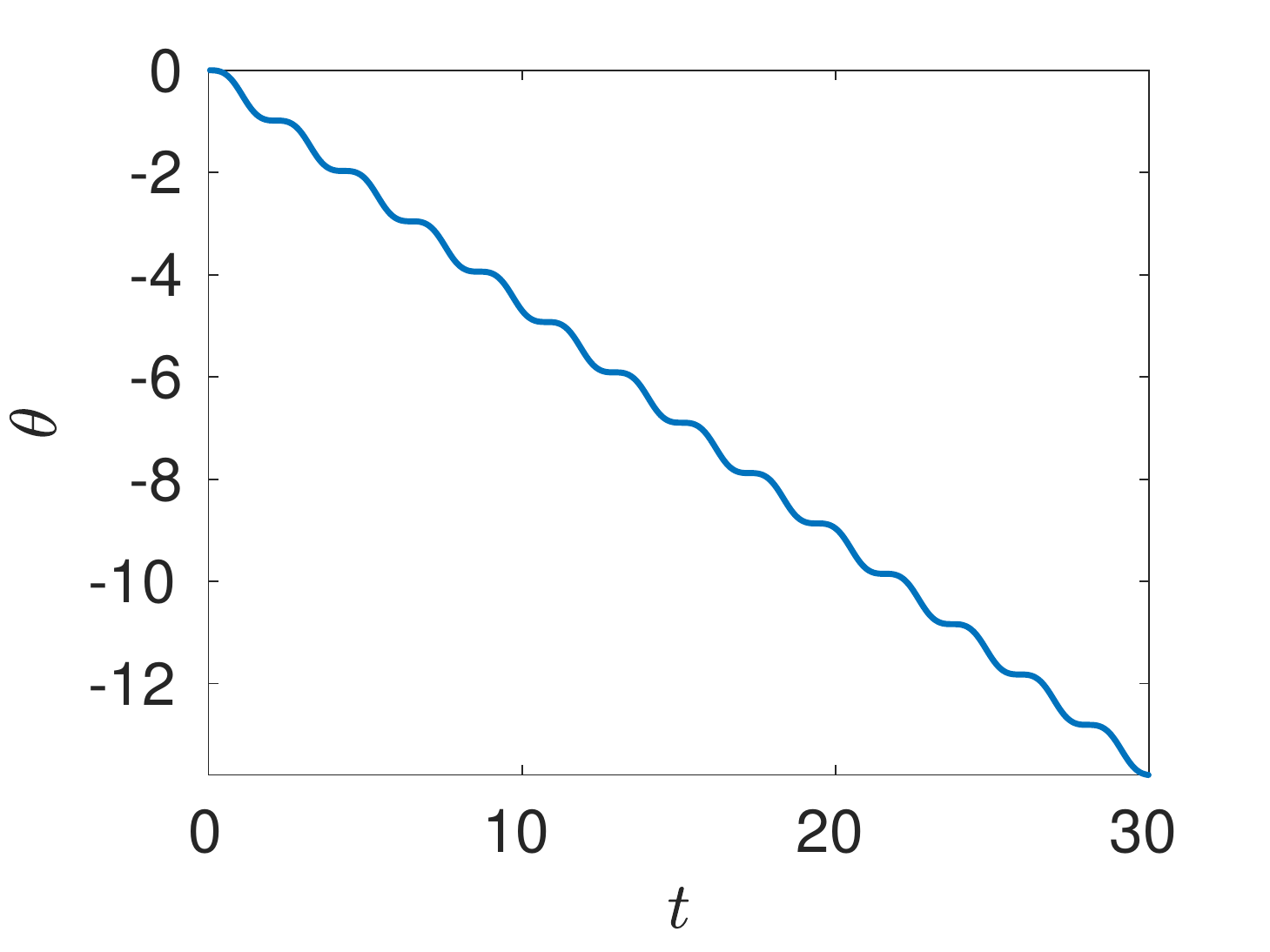}
\includegraphics[width=0.49\textwidth]{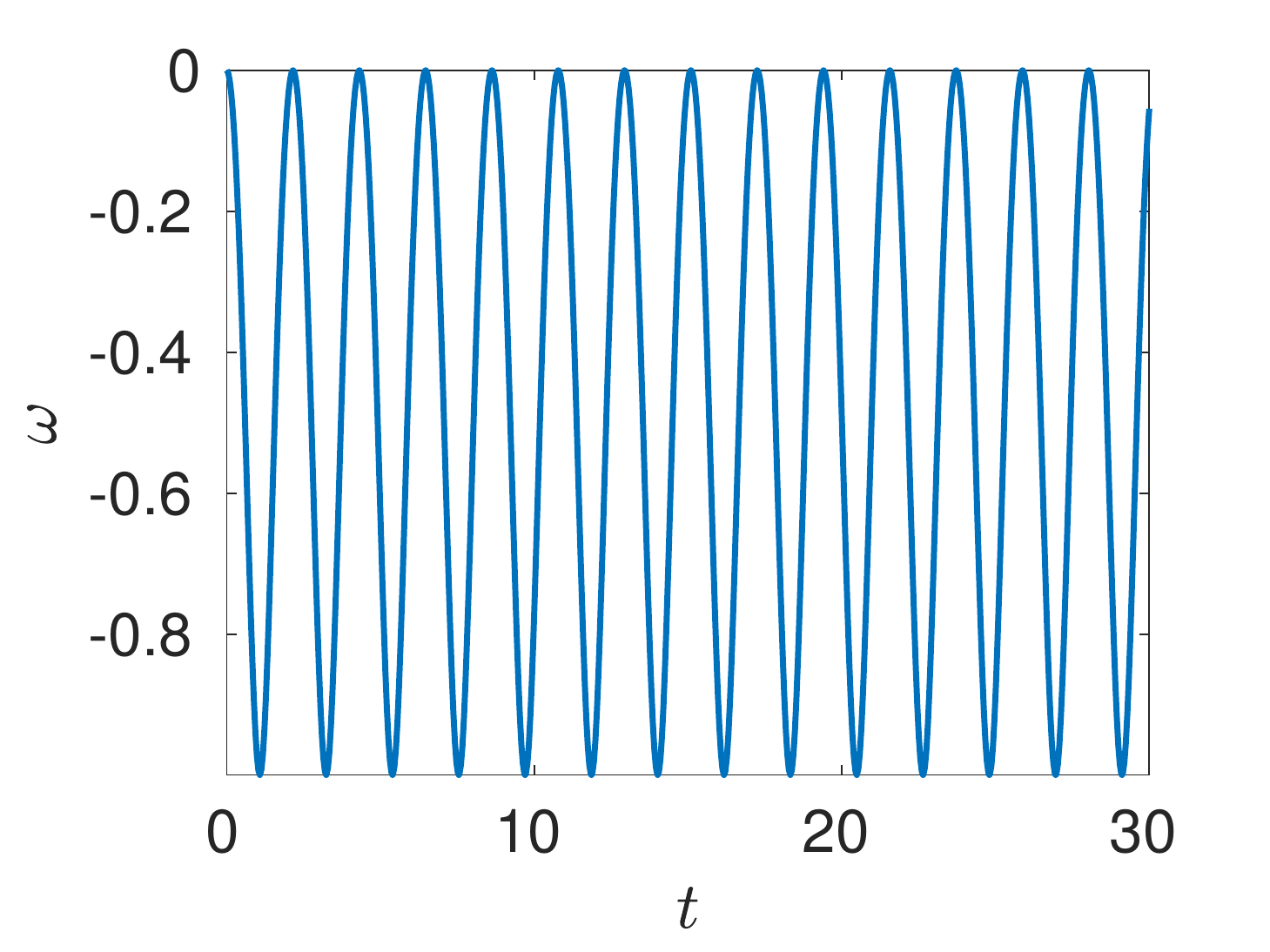}
\vspace{-0.15in}
$$
\text{(a)} \qquad\qquad\qquad\qquad\qquad\qquad\qquad\qquad \text{(b)}
$$
\includegraphics[width=0.49\textwidth]{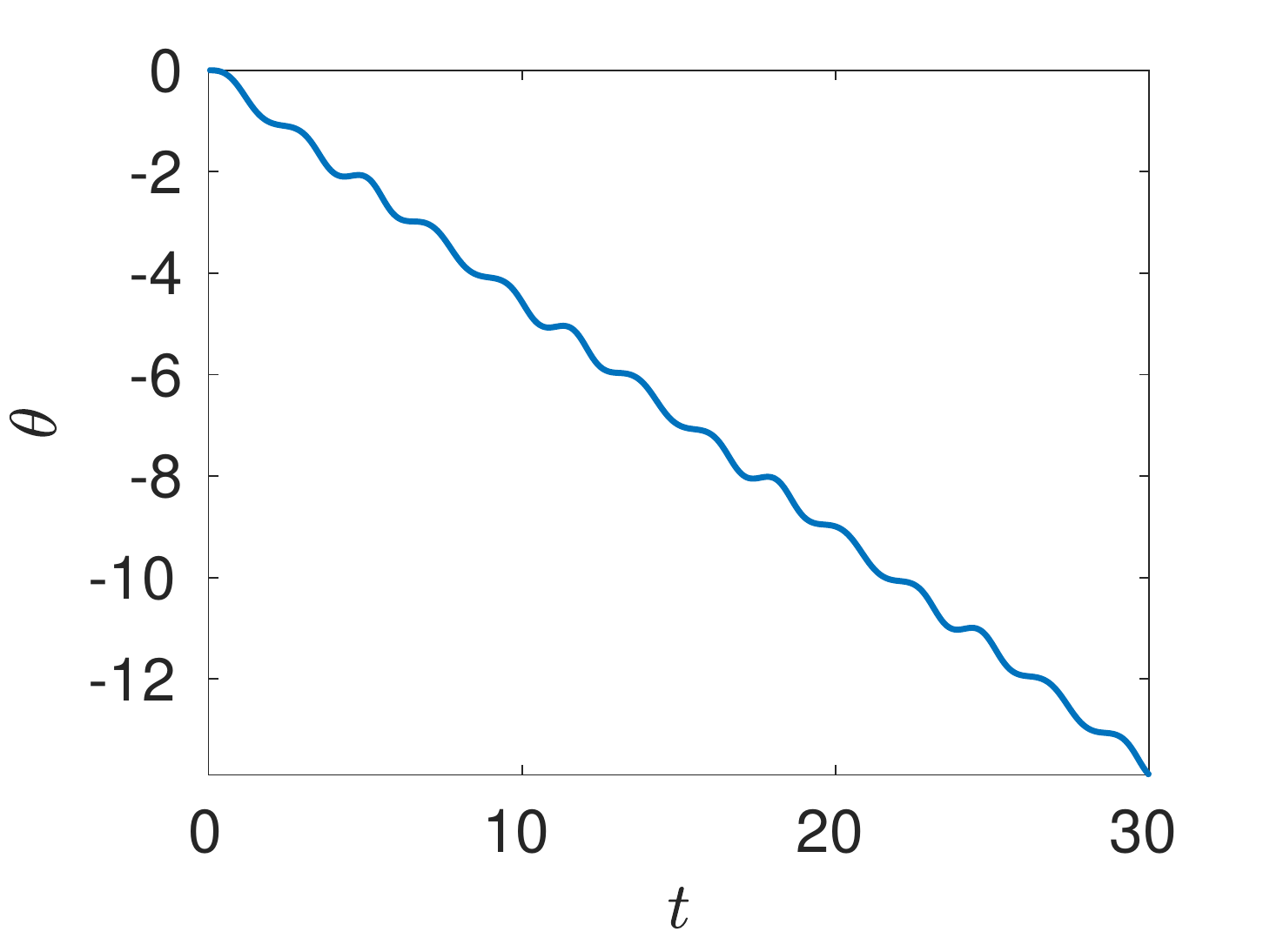}
\includegraphics[width=0.49\textwidth]{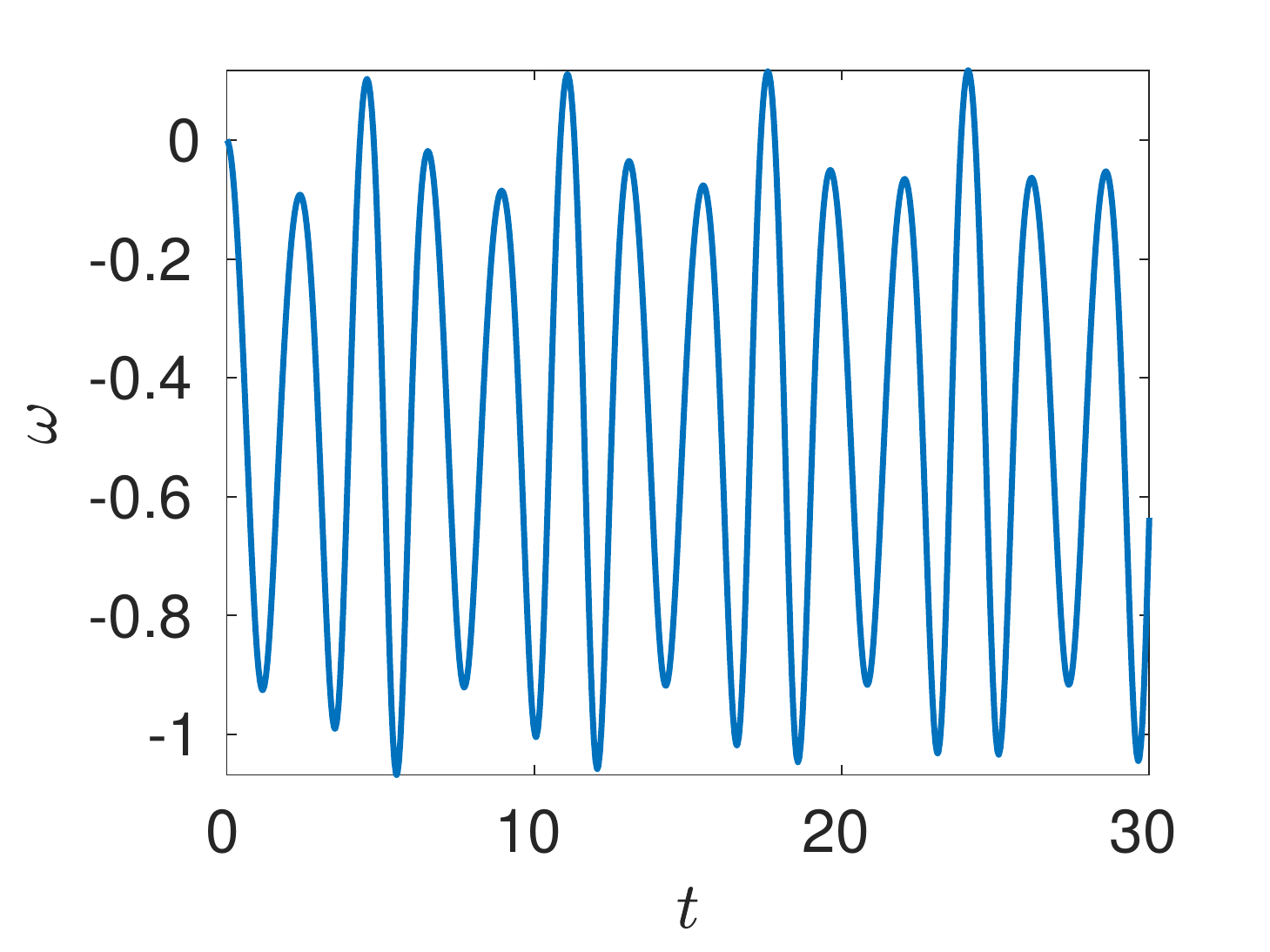}
\vspace{-0.15in}
$$
\text{(c)} \qquad\qquad\qquad\qquad\qquad\qquad\qquad\qquad \text{(d)}
$$
\includegraphics[width=0.49\textwidth]{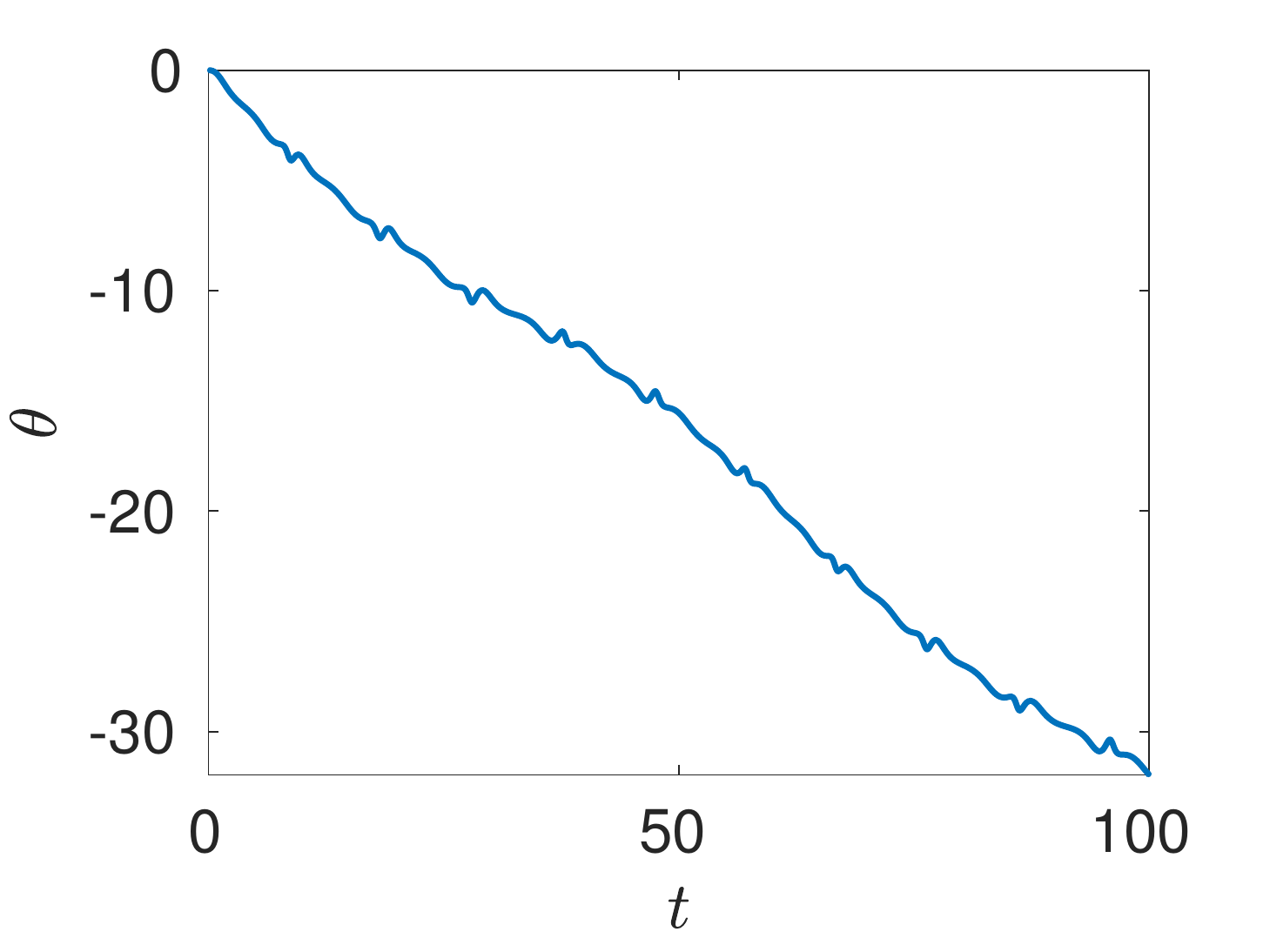}
\includegraphics[width=0.49\textwidth]{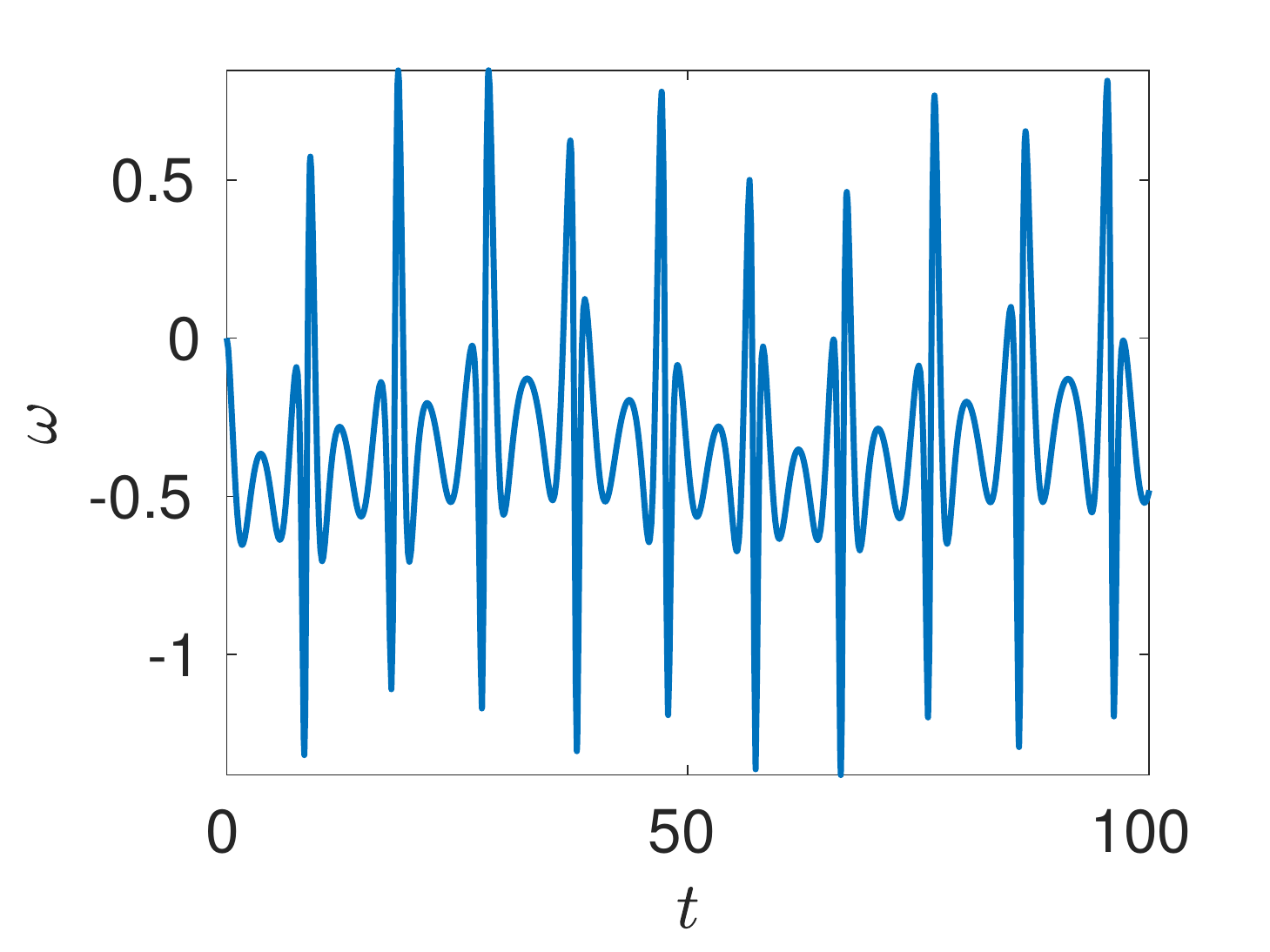}
\vspace{-0.15in}
$$
\text{(e)} \qquad\qquad\qquad\qquad\qquad\qquad\qquad\qquad \text{(f)}
$$
\caption{Time series for the perturbed circular orbit showing $\theta(t)$ and $\omega(t)=\dot{\theta}(t)$ with both initially set to $\theta(0) = \omega(0)=0$ for $\dot{y}_2(0) = 0$ (a)-(b), $\dot{y}_2(0) = 0.1$ (c)-(d), and $\dot{y}_2(0) = 0.5$ (e)-(f) in each case.}\label{TimeSeries2}
\end{figure}

In Fig. \ref{TimeSeries2} we plot some simulations of the rotation of the small body as it orbits around the larger one. We see that for a perfectly circular orbit, the body rotates primarily in one direction with some `wobble' due to the orbital forcing. The angular velocity reveals this to just be integrated periodic motion, and so it is also periodic (modulo the period). As the velocity perturbation is increased, multiple-frequency effects are observed in the angular velocity, and these correspond to variations in the wobbling previously seen at intermittent intervals. Finally for large enough perturbations, these small irregularities begin to occur more frequently, and the angular position of the body becomes increasingly unpredictable. These behaviours are generic throughout the portions of the phase space that we explored.

\begin{figure}
\begin{center}
\includegraphics[width=0.4\textwidth]{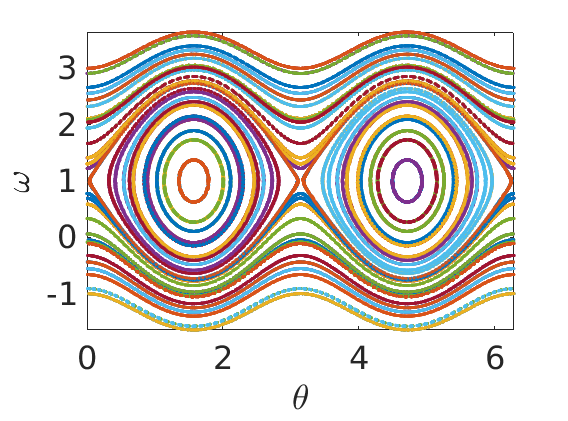}
\includegraphics[width=0.4\textwidth]{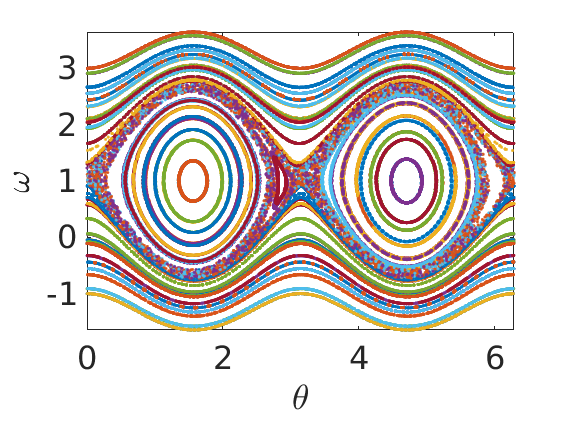}
\vspace{-0.15in}
$$
\text{(a)} \qquad\qquad\qquad\qquad\qquad\qquad\qquad\qquad \text{(b)}
$$
\includegraphics[width=0.4\textwidth]{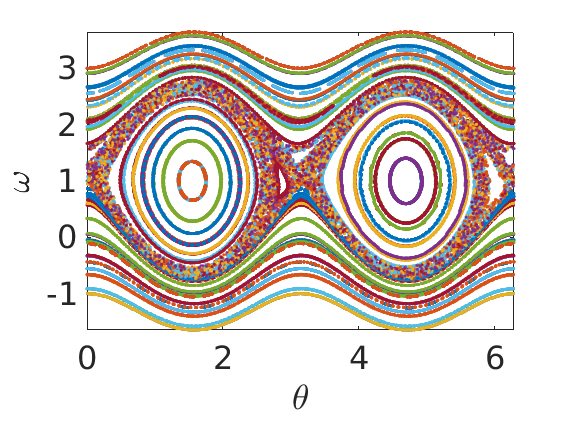}
\includegraphics[width=0.4\textwidth]{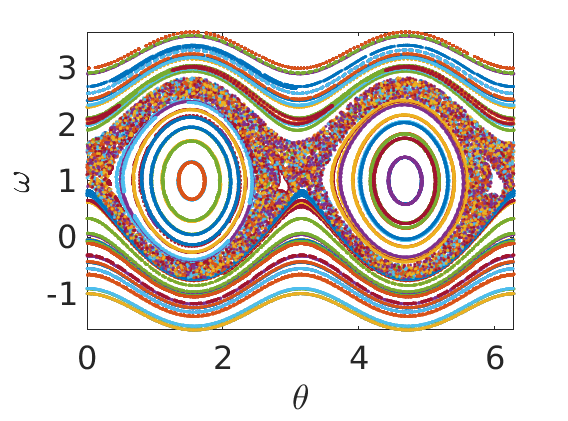}
\vspace{-0.15in}
$$
\text{(c)} \qquad\qquad\qquad\qquad\qquad\qquad\qquad\qquad \text{(d)}
$$
\includegraphics[width=0.4\textwidth]{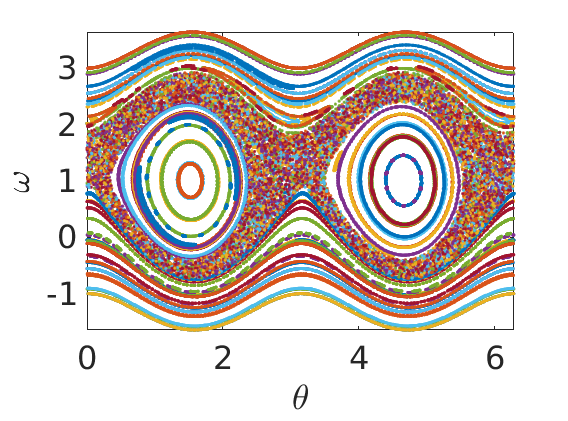}
\includegraphics[width=0.4\textwidth]{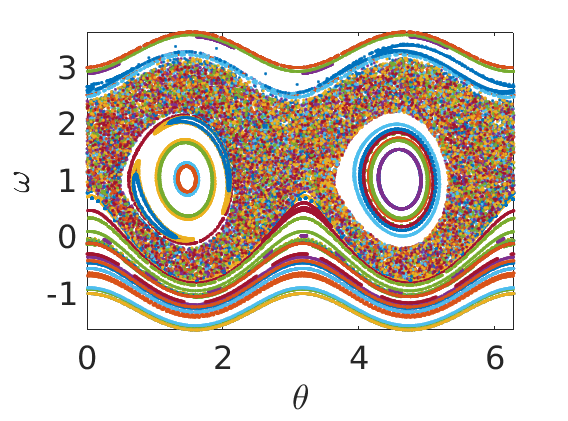}
\vspace{-0.15in}
$$
\text{(e)} \qquad\qquad\qquad\qquad\qquad\qquad\qquad\qquad \text{(f)}
$$
\includegraphics[width=0.4\textwidth]{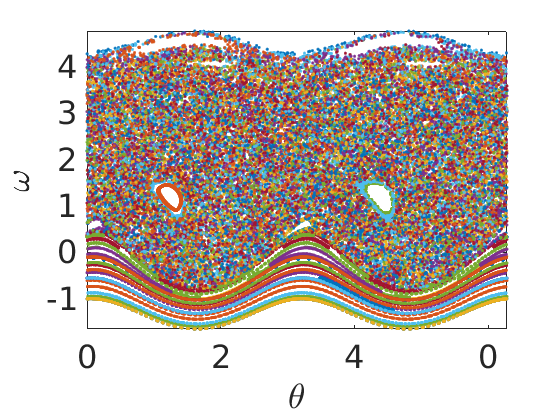}
\includegraphics[width=0.4\textwidth]{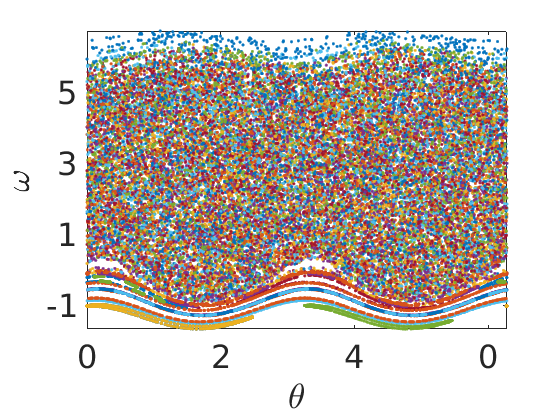}
\vspace{-0.15in}
$$
\text{(g)} \qquad\qquad\qquad\qquad\qquad\qquad\qquad\qquad \text{(h)}
$$
\caption{Poincar\'{e} sections over $\theta$ and $\omega = \dot{\theta}$ with the bifurcation parameter $\dot{y}_2(0)$ taking the values $\dot{y}_2(0) = 0$ (a), $\dot{y}_2(0) =0.01$ (b), $\dot{y}_2(0) = 0.02$ (c), $\dot{y}_2(0) = 0.03$ (d), $\dot{y}_2(0) = 0.05$ (e), $\dot{y}_2(0) = 0.1$ (f), $\dot{y}_2(0) = 0.3$ (g), and $\dot{y}_2(0) = 0.5$ (h).}\label{v}
\end{center}
\end{figure}

We construct a Poincar\'{e} section for this system by considering the times $T_p = \{t \in \mathbb{R} : x_2(t) = 0, \dot{x}_2(t)>0 \}$ corresponding to the bottom of the orbit in Fig. \ref{Orbits}. We vary the initial angle $\theta(0)$ from $0$ to $2\pi$ in $10$ increments, and the initial angular velocities $\dot{\theta}(0) = \omega(0)$ from $-1$ to $3$ in $10$ increments, and plot values of $\omega(t)$ against $\theta(t)\mod 2 \pi$ for $t \in T_p$ in Fig. \ref{v}.

For no perturbation, we see only periodic orbits, corresponding to rotation that `wobbles' in its orbit with a periodic motion. As we perturb the orbit slightly, we see that specific orbits are destroyed due to resonance with the slightly-perturbed orbit. Further increasing the perturbation, we see a transition to global chaotic dynamics throughout the phase space, with small islands of stability remaining by the final plot \cite{regchaos}.

\subsection{Simulations for perturbed figure-8 orbits}
We now consider the three-body problem with a figure-8 orbit as shown in Fig. \ref{Orbits}. All three bodies have planar motion along the figure-8 curve. See \cite{chenciner2000remarkable,chenciner2005rotating,nauenberg2007continuity} for further discussion on figure-8 orbits. Given that the orbital dynamics specified for the rigid body being a 1D rod is given by the point mass approximation to leading order, with first order corrections vanishing and second order corrections being negligible in the regime of the size of the rigid body being much smaller compared to the separation of bodies in the system, the figure-8 orbit is stable as discussed by work in \cite{HuSun2009,Roberts2007}. For instances where the rigid body can be modelled via perturbations to a sphere and for trajectories that approach on a length scale that is comparable in size to the geometry of the rigid body, such as for the models considered by \cite{Correia2014,Delisle2017}, higher order terms would cause this particular periodic orbit to become unstable. 

This orbit is far from circular, and hence the analytical chaos results for perturbations of circular orbits presented in the previous section do not hold here. However, we shall still demonstrate via numerical simulations that chaos is ubiquitous in this configuration, suggesting that the circular orbit is the special case where chaos does not exist.

\begin{figure}
\includegraphics[width=0.49\textwidth]{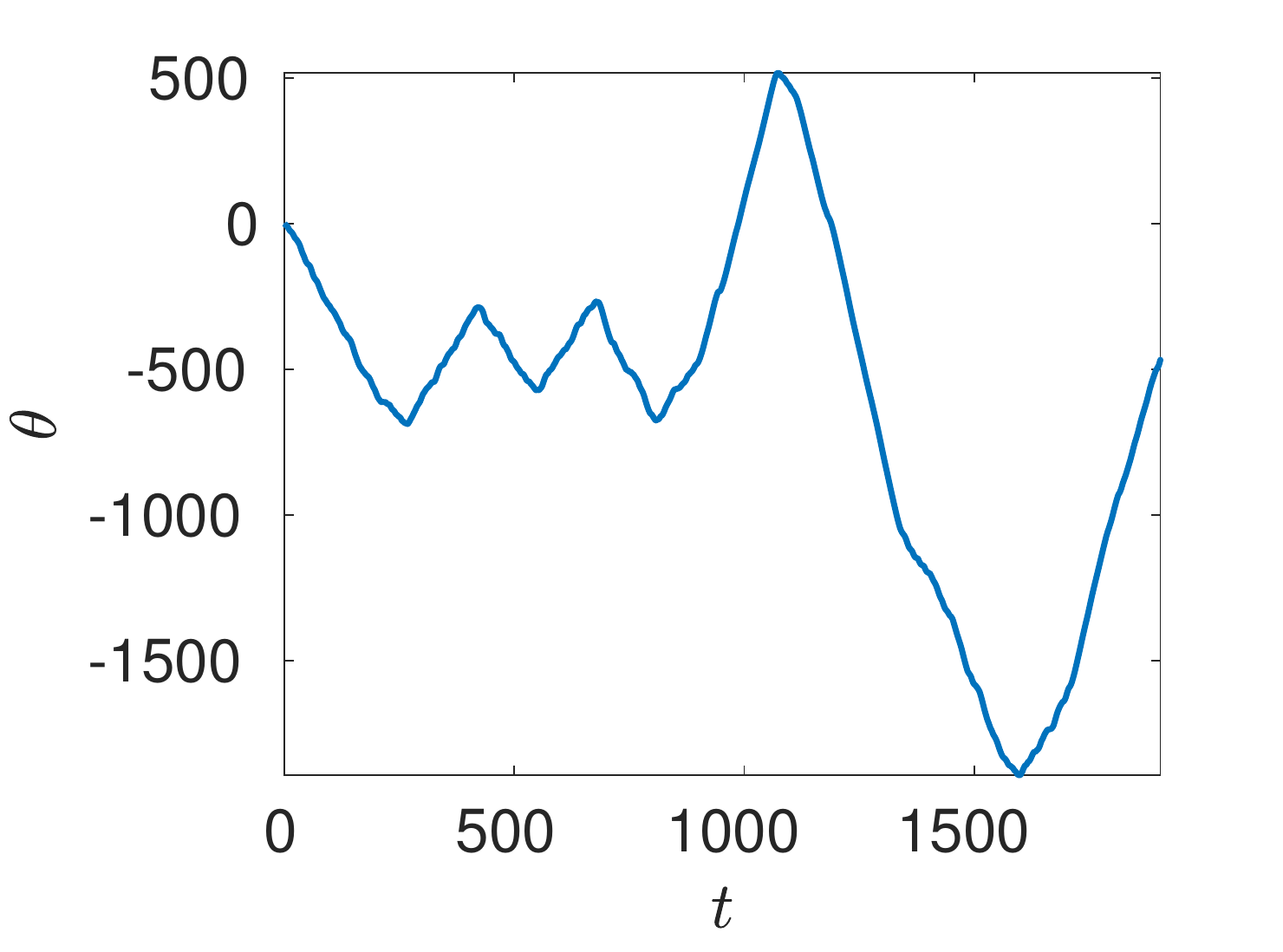}
\includegraphics[width=0.49\textwidth]{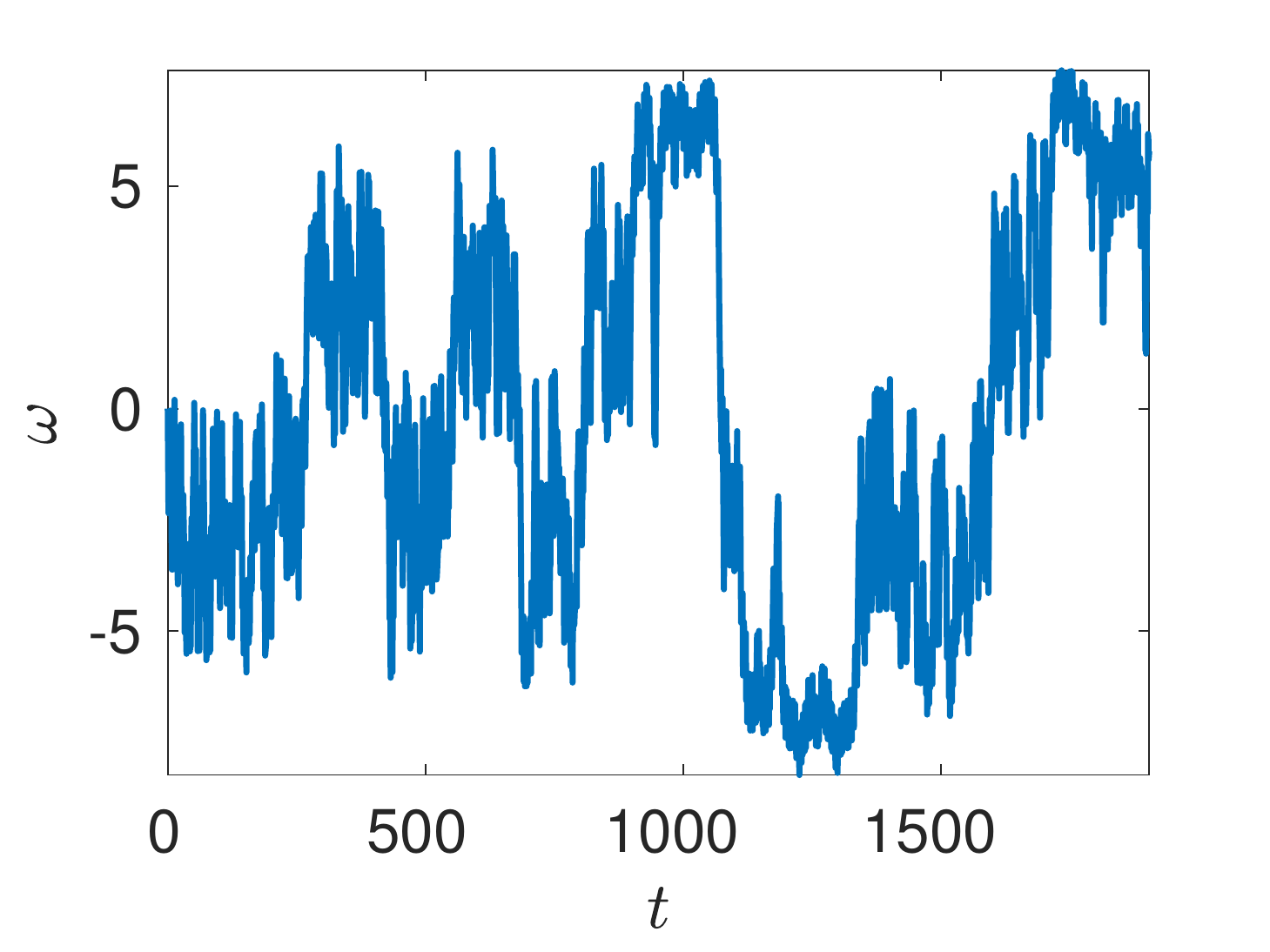}
\vspace{-0.15in}
$$
\text{(a)} \qquad\qquad\qquad\qquad\qquad\qquad\qquad\qquad \text{(b)}
$$
\includegraphics[width=0.49\textwidth]{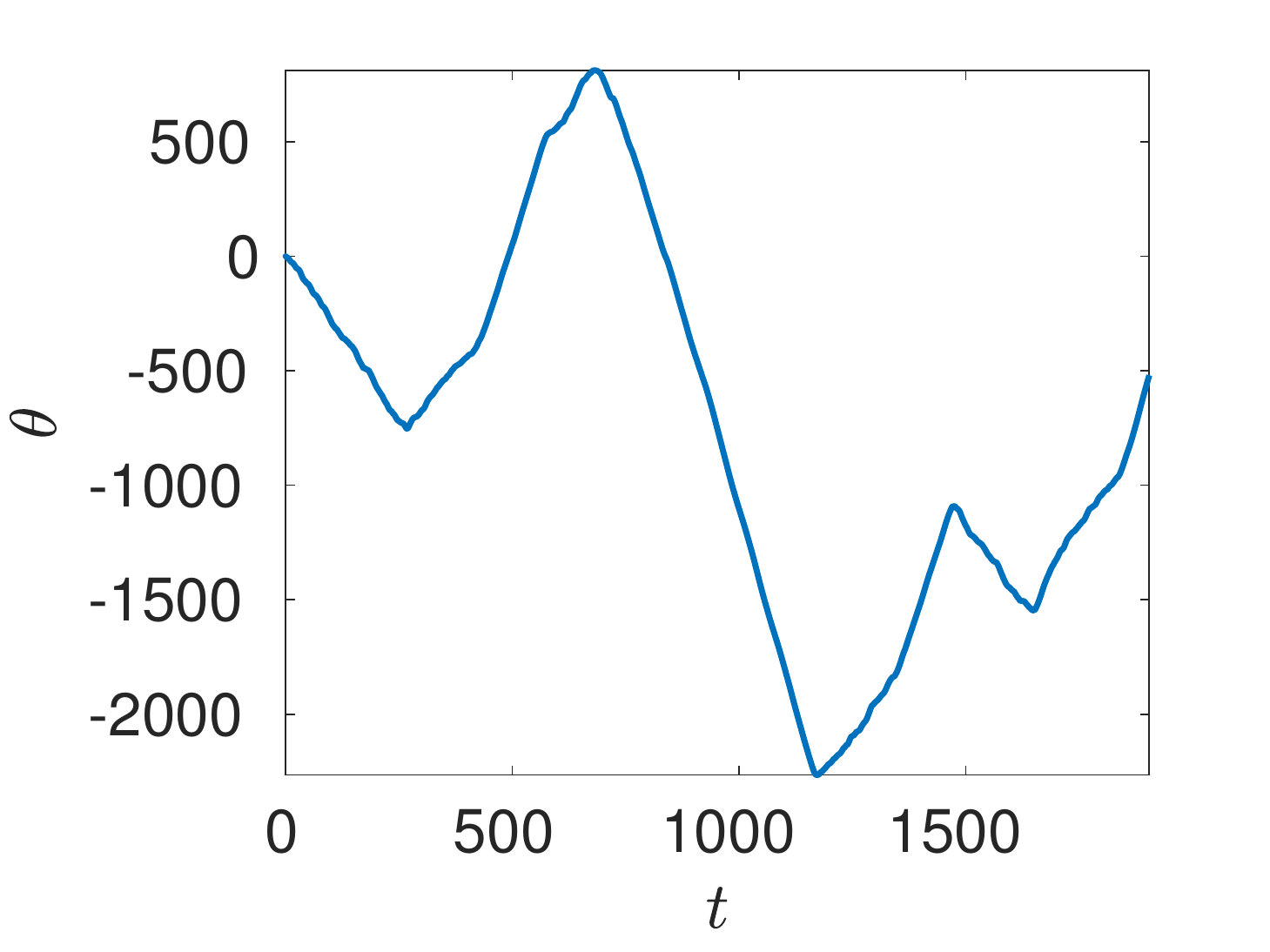}
\includegraphics[width=0.49\textwidth]{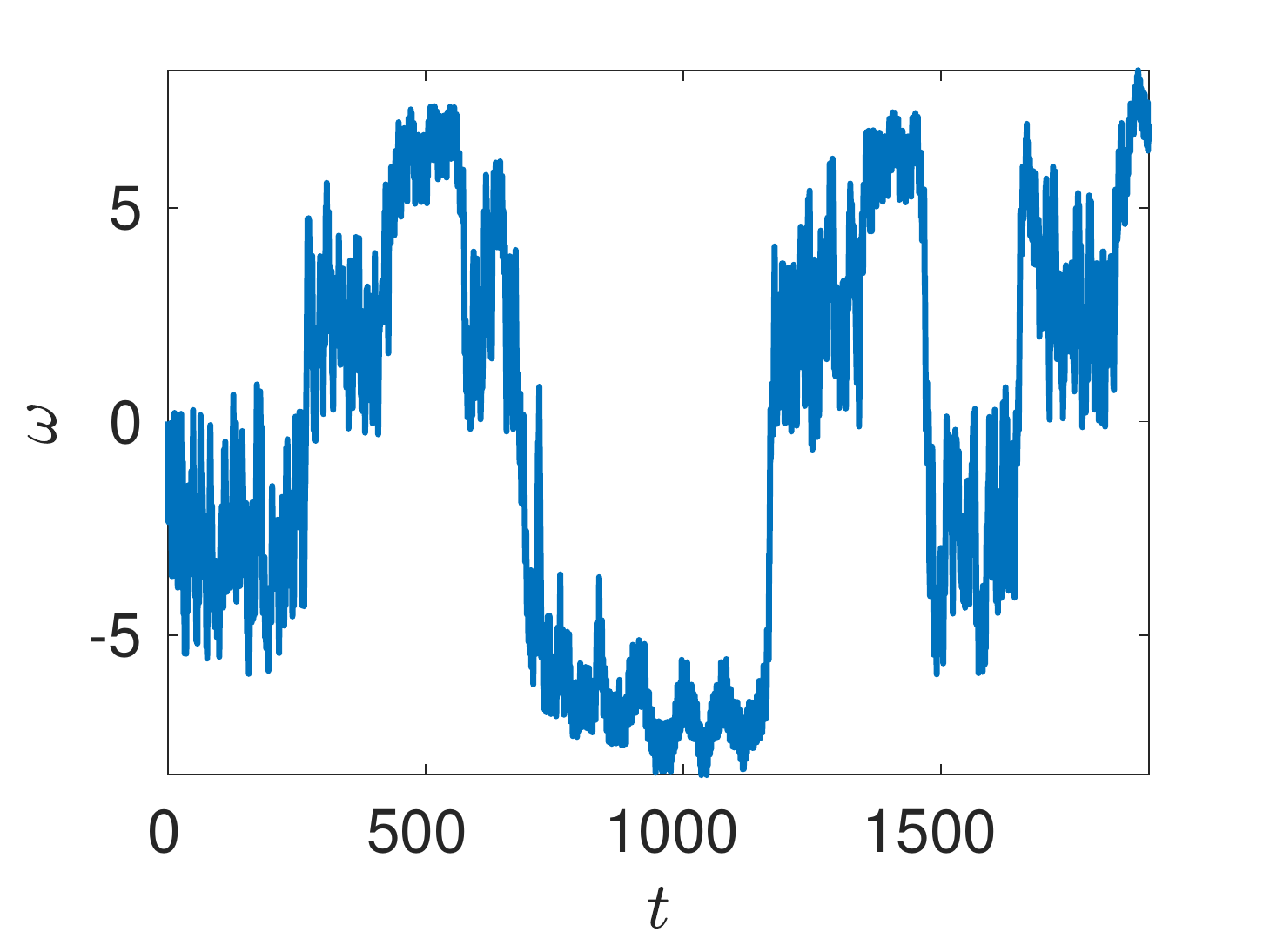}
\vspace{-0.15in}
$$
\text{(c)} \qquad\qquad\qquad\qquad\qquad\qquad\qquad\qquad \text{(d)}
$$
\includegraphics[width=0.49\textwidth]{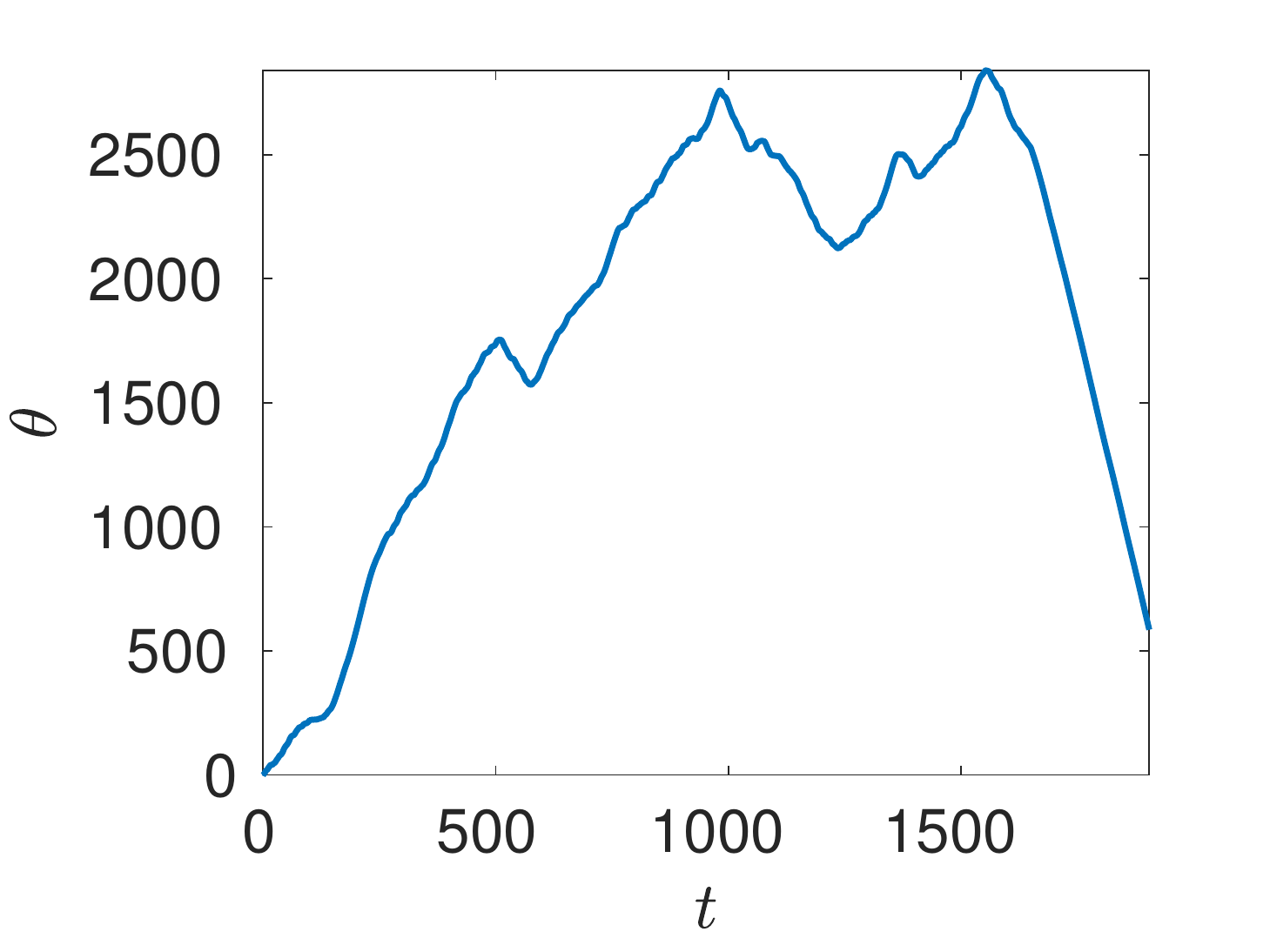}
\includegraphics[width=0.49\textwidth]{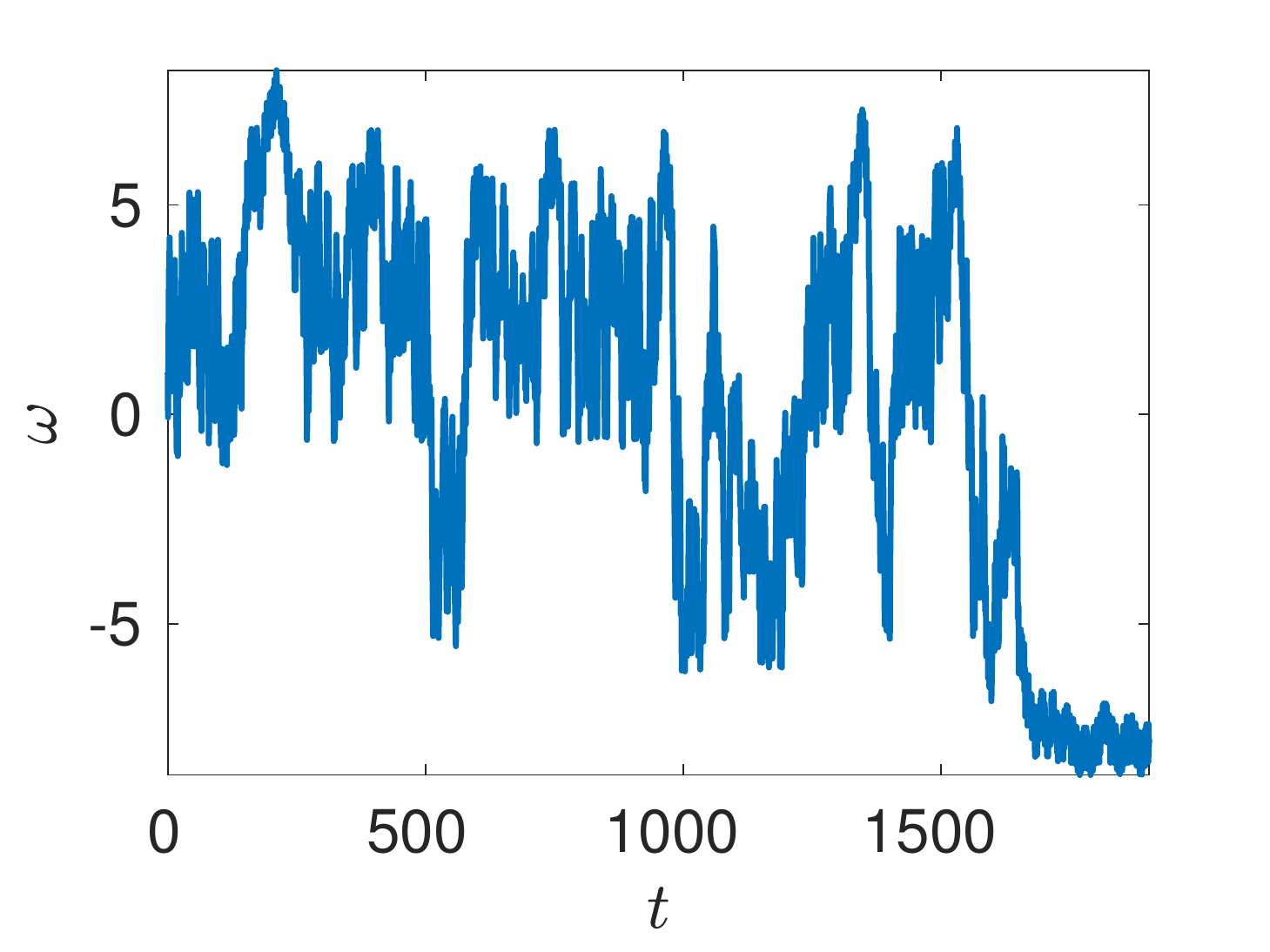}
\vspace{-0.15in}
$$
\text{(e)} \qquad\qquad\qquad\qquad\qquad\qquad\qquad\qquad \text{(f)}
$$
\caption{Time series for the figure-8 orbit showing $\theta(t)$ and $\omega(t)=\dot{\theta}(t)$ with $\theta(0)=0$, for $\omega(0) = 0$ (a)-(b), $\omega(0) = 10^{-5}$ (c)-(d), and $\omega(0) = 1$ (e)-(f).}\label{TimeSeries3}
\end{figure}

\begin{figure}
\begin{center}
\includegraphics[width=.4\textwidth]{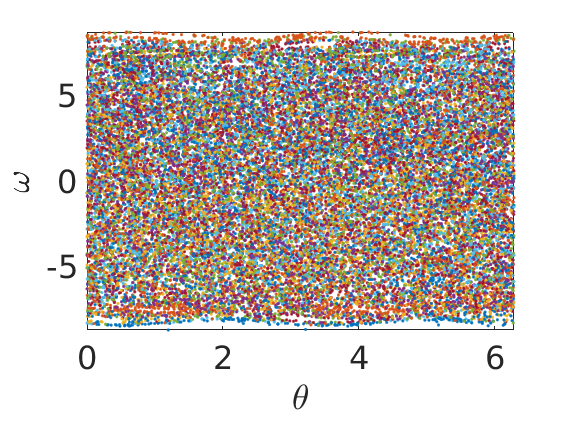}
\vspace{-0.15in}
\end{center}
\caption{A Poincar\'{e} section over $\theta$ and $\omega = \dot{\theta}$ for the figure-8 orbit in Figure \ref{Orbits}.}\label{I}
\end{figure}

In Fig. \ref{TimeSeries3}, we plot the angle and angular velocity corresponding to three different values of $\omega(0)$. Both angular position and velocity exhibit chaotic motions. In particular the angular position $\theta(t)$ has a run-and-tumble behaviour where the body is spinning in one direction for many orbits before it begins rotating in the opposite direction (note that for this simulation, the three-body orbit has a period $T \approx 6.3490473929$, as reported in \cite{li2017one}). 

In Fig. \ref{I} we plot a Poincar\'{e} section for $\omega(0)=0$ using the same discretization of initial values as before, and the same definition of the Poincar\'{e} section at $T_p$ for $300$ orbits. As expected, we see what appears to be completely ergodic behaviour indicative of global Hamiltonian chaos \cite{regchaos}.

The figure-8 orbits are but one example of the possible orbits for the planar three-body problem, and therefore constitute another example where chaotic dynamics emerge in the rotational motion of the rigid bodies. We found similar results for many of the planar three-body problem orbits, recently discovered in \cite{li2017one,dmitravsinovic2017newtonian}. This suggests that such chaos should be ubiquitous in the $N$-body problem where one body is non-spherical.

\section{Discussion}

We have considered a gravitational $N$-body system where one of the bodies is a rotating rigid body of arbitrary geometry, which generalizes many models that only consider point masses or purely spherical bodies. In order to better understand these dynamics, we study a reduced model consisting of a planar $N$-body problem where the rigid body is taken to have a 1D geometry. In particular, the body is treated as a rod with an arbitrary mass distribution. The key feature to note is that the rotation of a more complicated 3D structure can be approximated by a 1D structure when rigid body motions are within a plane. This simplification therefore makes the system more amenable to analysis yet still preserves key features of the full 3D structure of the rigid body. Note however that not all purely 1D systems which are stable necessarily have stable analogues for full 3D rotations, rather, if one can show that chaotic rotations occur for the 1D case, then this typically implies chaotic rotations occur in higher dimensions given that more degrees of freedom are introduced. 

The translational motion equations are equivalent to the $N$-body problem for point masses on co-planar orbits, while the additional rotational motion equation, arising from asymptotic corrections to the gravitational force, is included for the rigid body. This extra equation becomes trivial only when the gravitational force acting on the rod is zero, regardless of the mass distribution of the body. The translational equations of motions do not depend on the rotational motion in the point-particle approximation, and thus they can be solved just as for $N$-body problems in order to obtain the orbits. Once the orbits are determined, one can place the rotational motion equation on top of these results, feeding the orbits as a non-autonomous forcing. 

We analytically demonstrate the existence of homoclinic chaos in the case where one of the orbits is nearly circular by way of the Melnikov method, and give numerical evidence for such chaos when the orbits are more complicated (such as more extreme elliptical orbits or figure-8 orbits). These results suggest that chaos is ubiquitous in such $N$-body problems when one or more of the rigid bodies is non-spherical. We show that the extent of chaos in parameter space is strongly tied to the deviations from purely circular orbits. Such dynamics give a possible explanation for routes to chaotic dynamics observed in many-body systems such as the Pluto system in which some of the bodies are better approximated as rods rather than perfect spheres.

The circular orbits and figure-8 orbits are specific examples, and due to the wide appearance of chaos in the parameter space of these problems, we can likely find similar results for many of the planar three-body problem orbits, such as those recently discovered in \cite{dmitravsinovic2017newtonian,li2017one}. More generally, since all that is required for the appearance of chaos is for one rigid body to exhibit chaotic rotations, the results naturally extend to $N$-bodies, in planar or non-planar (fully 3D) configurations. This suggests that such chaos should be ubiquitous in the $N$-body problem when one or more of the bodies are rigid and non-spherical.  

Returning to the motivating application, recall that in the Pluto system, the asymmetric Styx does appear to exhibit intermittent obliquity variations and episodes of tumbling, suggesting some form of chaos in the rotational dynamics \cite{showalter2015resonant,correia2015spin,quillen2017obliquity}. The results we obtain suggest that chaos could be ubiquitous in the rotational dynamics of such small moons when they are geometrically asymmetric, given its existence in the absence of tidal dissipation, obliquity variation, and orbital resonance. This outcome could explain why Styx appears to exhibit rotational chaos or tumbling while other satellites or moons in the Pluto system, which are more symmetric, may not. Indeed, the emergence of such chaotic dynamics from only orbital forcing and non-sphericity of the rigid body suggests that such chaotic tumbling may be prevalent in many systems where some of the bodies have asymmetric geometries.

\subsection*{Acknowledgments}
The authors would like to thank A. Goriely for helpful and worthwhile discussions and T. G. Bollea for inspiration. 

\section{Appendix: Derivation of orbital mechanics for a single rigid rotating rod}

We first give a general formulation of the problem of $N$ orbiting bodies, with $N - 1$ point masses and a single rigid body of arbitrary shape, which we will refer to as the $N$th body hereafter. We assume that the bodies are separated on a length scale much larger than the geometry of the body. We then restrict our attention to a 1D geometry for the rigid body, resulting in a single rotational equation for a rod of arbitrary mass distribution.

\subsection{Rotating rigid body under gravitational forces}

We study a system of $N$ rigid bodies that exert gravitational forces on each other. The $N$th body has a reference configuration $\mathcal{B}_{0}\subset\mathbb{R}^{3}$ initially and is mapped to a current configuration $\mathcal{B}_{t}\subset\mathbb{R}^{3}$ at time $t$. The position vector $\boldsymbol{r}_{N}$ points from an arbitrary origin to the body's center of mass. For a rigid body, the center of mass occurs at the same material point in the reference configuration, however the body itself will translate and rotate according to the effects of gravity. 

We determine the time-evolution of the translation of the point masses. In particular,  we make use of the balance of linear momentum to obtain that the acceleration of the $i$th point mass at time $t$ is
\begin{equation}
m_{i}\frac{\mathrm{d}\boldsymbol{v}_{i}}{\mathrm{d}t}=\sum_{j\neq i}^{N-1}\frac{Gm_{i}m_{j}}{\left|\boldsymbol{r}_{j}-\boldsymbol{r}_{i}\right|^{3}}\left(\boldsymbol{r}_{j}-\boldsymbol{r}_{i}\right)+Gm_{i}\intop_{\mathcal{B}_{t}}\frac{\rho\left(\boldsymbol{R}_{t}\right)}{\left|\boldsymbol{R}_{t}-\boldsymbol{r}_{i}\right|^{3}}\left(\boldsymbol{R}_{t}-\boldsymbol{r}_{i}\right)\mathrm{d}V_{t}, \label{eq:BLMPointmass}
\end{equation}
where the integral is taken over the volume in the current configuration $\mathcal{B}_{t}$, $\rho\left(\boldsymbol{R}_{t}\right)$ is the mass density of the $N$th continuum, $\boldsymbol{R}_{t}$ is the position vector of a
material point in $\mathcal{B}_{t}$, $G$ is the universal gravitational constant, $\frac{\mathrm{d}\boldsymbol{v}_{i}}{\mathrm{d}t}$ is the acceleration field of the $i$th body, and $m_j$ for $j=1\dots N-1$ are the masses of the point masses. The mass of the $N$th body is constant for all time and is given as
\begin{equation}
m_{N}=\intop_{\mathcal{B}_{t}}\rho\left(\boldsymbol{R}_{t}\right)\mathrm{d}V_{t}.\label{eq:Mass}
\end{equation}

We note that the first term in \eqref{eq:BLMPointmass} is due to the gravitational effects of other point masses in the system and the second term are the gravitational effects of the body due to its finite geometry. To determine the integrals that arise from the second term, we introduce two vectors: The separation between the $i$th and $N$th center of masses $\boldsymbol{\xi}_{i} = \boldsymbol{r}_{N} - \boldsymbol{r}_{i}$ and the displacement between any material point in the $N$th body to its center of mass $\boldsymbol{d} = \boldsymbol{r}_{N} - \boldsymbol{R}_{t}$. We suppose that the separation between the bodies is much larger than the finite geometry of the $N$th body, so that
\begin{equation}
\frac{\left|\boldsymbol{d}\right|}{\left|\boldsymbol{\xi}_{i}\right|} = \varepsilon_{i} \ll 1. \label{eq:Pointmassasy}
\end{equation}

Under this asymptotic regime, the integral term on the right hand side of \eqref{eq:BLMPointmass} can be expanded as a quadrupole expansion to give
\begin{equation}
Gm_{i}\intop_{\mathcal{B}_{t}}\frac{\rho\left(\boldsymbol{R}_{t}\right) \left(\boldsymbol{\xi}_{i}-\boldsymbol{d}\right)}{\left|\boldsymbol{\xi}_{i}-\boldsymbol{d}\right|^{3}}
\mathrm{d}V_{t}
=Gm_{i}\intop_{\mathcal{B}_{t}}\frac{\rho\left(\boldsymbol{R}_{t}\right)}{\left|\boldsymbol{\xi}_{i}\right|^{2}}\left(\hat{\boldsymbol{\xi}_{i}}-\varepsilon_{i}\hat{\boldsymbol{d}}+3\varepsilon_{i}\left(\hat{\boldsymbol{\xi}_{i}}\cdot\hat{\boldsymbol{d}}\right)\hat{\boldsymbol{\xi}_{i}}+O\left(\varepsilon_{i}^{2}\right)\right)\mathrm{d}V_{t},
\end{equation}
where $\hat{\square}$ is the corresponding unit vector, so that to first order in $\varepsilon_{i}$, \eqref{eq:BLMPointmass} gives
\begin{multline}
\frac{\mathrm{d}\boldsymbol{v}_{i}}{\mathrm{d}t}\sim\sum_{j\neq i}^{N}\frac{Gm_{j}}{\left|\boldsymbol{r}_{j}-\boldsymbol{r}_{i}\right|^{3}}\left(\boldsymbol{r}_{j}-\boldsymbol{r}_{i}\right)\\
+G\intop_{\mathcal{B}_{t}}\frac{\rho\left(\boldsymbol{R}_{t}\right)}{\left|\boldsymbol{r}_{N}-\boldsymbol{r}_{i}\right|^{3}}\left[\left(\boldsymbol{R}_{t}-\boldsymbol{r}_{N}\right)+\frac{3\left(\boldsymbol{r}_{N}-\boldsymbol{r}_{i}\right)\cdot\left(\boldsymbol{r}_{N}-\boldsymbol{R}_{t}\right)}{\left|\boldsymbol{r}_{N}-\boldsymbol{r}_{i}\right|^{2}}\left(\boldsymbol{r}_{N}-\boldsymbol{r}_{i}\right)\right]\mathrm{d}V_{t}, \label{eq:BLMPointmassFinal}
\end{multline}
in the limit of \eqref{eq:Pointmassasy} and where we have used (\ref{eq:Mass}) for the $N$th body. Note that we have also absorbed the point mass approximation of the rigid body into the summation.

\begin{figure}
\begin{centering}
\includegraphics[scale=0.3]{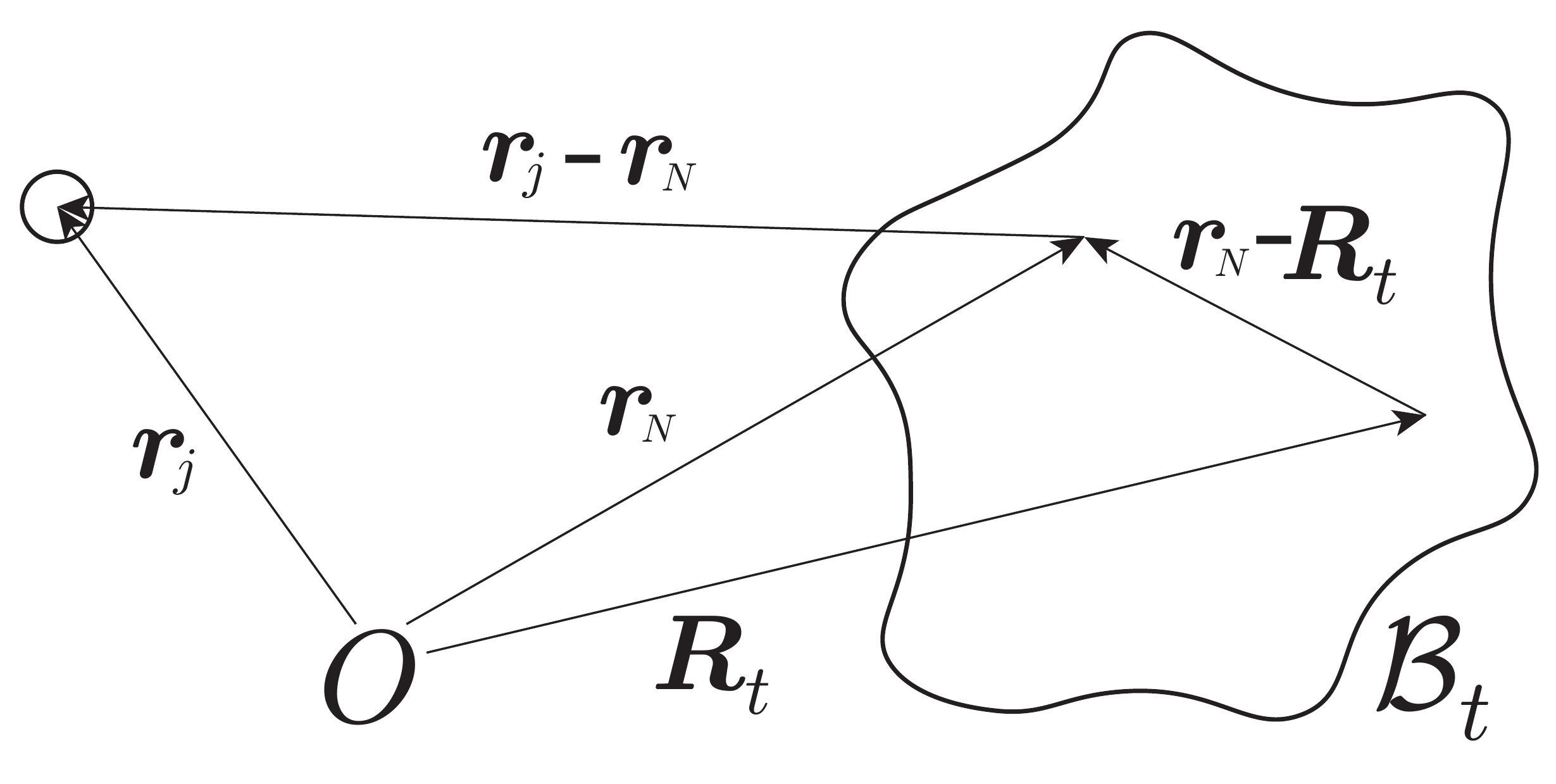}
\par\end{centering}
\caption{The kinematic description of the $N$ rigid body system. The $N$th
continuum in the current configuration $\mathcal{B}_{t}$
at time $t$ has a center of mass at $\boldsymbol{r}_{N}$ from an
arbitrary origin $O$, whilst all other material points have positions
denoted by $\boldsymbol{R}_{t}$.\label{fig:GeneralRigid}}
\end{figure}

Similarly, for the $N$th body with finite geometry, the balance of linear
momentum at time $t$ is (see Fig.~\ref{fig:GeneralRigid})
\begin{equation}
 \intop_{\mathcal{B}_{t}}\rho\left(\boldsymbol{R}_{t}\right)\frac{\mathrm{d}\boldsymbol{v}_{N}}{\mathrm{d}t}\mathrm{d}V_{t}
 =\sum_{j=1}^{N-1}Gm_{j}\intop_{\mathcal{B}_{t}}\frac{\rho\left(\boldsymbol{R}_{t}\right)}{\left|\boldsymbol{r}_{j}-\boldsymbol{R}_{t}\right|^{3}}
\left(\boldsymbol{r}_{j}-\boldsymbol{R}_{t}\right)
\mathrm{d}V_{t}.\label{eq:BLM}
\end{equation}

As before, we introduce a vector describing the separation between the $j$th and $N$th center of masses $\boldsymbol{\xi}_{j} = \boldsymbol{r}_{j} - \boldsymbol{r}_{N}$ and another vector describing geometry of the $N$th body from its center of mass $\boldsymbol{d} = \boldsymbol{r}_{N} - \boldsymbol{R}_{t}$ and consider the asymptotic limit of
\begin{equation}
\frac{\left|\boldsymbol{d}\right|}{\left|\boldsymbol{\xi}_{j}\right|} = \varepsilon_{j} \ll 1.\label{eq:Rigidbodyasy}
\end{equation}

In this case, \eqref{eq:BLM} becomes
\begin{multline}
\intop_{\mathcal{B}_{t}}\rho\left(\boldsymbol{R}_{t}\right)\frac{\mathrm{d}\boldsymbol{v}_{N}}{\mathrm{d}t}\mathrm{d}V_{t}=\sum_{j=1}^{N-1}\frac{Gm_{j}m_{N}}{\left|\boldsymbol{r}_{j}-\boldsymbol{r}_{N}\right|^{3}}\left(\boldsymbol{r}_{j}-\boldsymbol{r}_{N}\right)\\
+Gm_{j}\intop_{\mathcal{B}_{t}}\frac{\rho\left(\boldsymbol{R}_{t}\right)}{\left|\boldsymbol{r}_{j}-\boldsymbol{r}_{N}\right|^{3}}\left[\left(\boldsymbol{r}_{N}-\boldsymbol{R}_{t}\right)-\frac{3\left(\boldsymbol{r}_{j}-\boldsymbol{r}_{N}\right)\cdot\left(\boldsymbol{r}_{N}-\boldsymbol{R}_{t}\right)}{\left|\boldsymbol{r}_{j}-\boldsymbol{r}_{N}\right|^{2}}\left(\boldsymbol{r}_{j}-\boldsymbol{r}_{N}\right)\right]\mathrm{d}V_{t}, \label{eq:BLMStep}
\end{multline}
where we have again used \eqref{eq:Mass}.

Additionally, for a rigid body whose translational acceleration is much larger than the rotational acceleration along an arbitrary axis (which is true provided that the rigid body is sufficiently far away from other gravitational bodies), $\frac{\mathrm{d}\boldsymbol{v}_{N}}{\mathrm{d}t}$ is approximately independent of the position coordinates $\boldsymbol{R}_{t}$, so that (\ref{eq:BLMStep}) reduces to
\begin{multline}
\frac{\mathrm{d}\boldsymbol{v}_{N}}{\mathrm{d}t}=\sum_{j=1}^{N-1}\frac{Gm_{j}}{\left|\boldsymbol{r}_{j}-\boldsymbol{r}_{N}\right|^{3}}\left(\boldsymbol{r}_{j}-\boldsymbol{r}_{N}\right)\\
+\frac{Gm_{j}}{m_{N}}\intop_{\mathcal{B}_{t}}\frac{\rho\left(\boldsymbol{R}_{t}\right)}{\left|\boldsymbol{r}_{j}-\boldsymbol{r}_{N}\right|^{3}}\left[\left(\boldsymbol{r}_{N}-\boldsymbol{R}_{t}\right)-\frac{3\left(\boldsymbol{r}_{j}-\boldsymbol{r}_{N}\right)\cdot\left(\boldsymbol{r}_{N}-\boldsymbol{R}_{t}\right)}{\left|\boldsymbol{r}_{j}-\boldsymbol{r}_{N}\right|^{2}}\left(\boldsymbol{r}_{j}-\boldsymbol{r}_{N}\right)\right]\mathrm{d}V_{t}.\label{eq:BLMFinal}
\end{multline}

To describe the time-evolution of the rigid body's rotation, we consider the balance of angular momentum in the current configuration 
\begin{equation}
\intop_{\mathcal{B}_{t}}\frac{\mathrm{d}}{\mathrm{d}t}\left[\left(\boldsymbol{R}_{t}-\boldsymbol{r}_{N}\right)\times\rho\left(\boldsymbol{R}_{t}\right)\boldsymbol{v}_{\mathrm{tan}}\right]\mathrm{d}V_{t}=\intop_{\mathcal{B}_{t}}\left(\boldsymbol{R}_{t}-\boldsymbol{r}_{N}\right)\times\boldsymbol{F}_{N}\mathrm{d}V_{t},\label{eq:BAM}
\end{equation}
where $\boldsymbol{v}_{\mathrm{tan}}$ is the tangential translational velocity as the body rotates around its center of mass and the force $\boldsymbol{F}_{N}$ is defined as,
\begin{equation}
\boldsymbol{F}_{N} = \sum_{j=1}^{N-1}\frac{Gm_{j}\rho\left(\boldsymbol{R}_{t}\right)}{\left|\boldsymbol{r}_{j}-\boldsymbol{R}_{t}\right|^{3}}\left(\boldsymbol{r}_{j}-\boldsymbol{R}_{t}\right).
\end{equation}

Introducing the separation vector $\boldsymbol{\xi}_{j}$ and rigid body vector $\boldsymbol{d}$, we expand the right hand side of \eqref{eq:BAM} in the limit specified in \eqref{eq:Rigidbodyasy} to find
\begin{multline}
-\sum_{j=1}^{N-1}Gm_{j}\intop_{\mathcal{B}_{t}}\boldsymbol{d}\times\rho\left(\boldsymbol{R}_{t}\right)\frac{\left(\boldsymbol{\xi}_{j}+\boldsymbol{d}\right)}{\left|\boldsymbol{\xi}_{j}+\boldsymbol{d}\right|^{3}}\mathrm{d}V_{t}\\
=-\sum_{j=1}^{N-1}Gm_{j}\intop_{\mathcal{B}_{t}}\boldsymbol{d}\times\frac{\rho\left(\boldsymbol{R}_{t}\right)}{\left|\boldsymbol{\xi}_{j}\right|^{2}}\left(\hat{\boldsymbol{\xi}_{j}}+\varepsilon_{j}\hat{\boldsymbol{d}}-3\varepsilon_{j}\left(\hat{\boldsymbol{\xi}_{j}}\cdot\hat{\boldsymbol{d}}\right)\hat{\boldsymbol{\xi}_{j}}+O\left(\varepsilon_{j}^{2}\right)\right)\mathrm{d}V_{t}.
\label{eq:BAM2}
\end{multline}

We note that since the cross product of a constant vector with respect to integration can be interpreted as a linear transform, we have that the first term on the right hand side of \eqref{eq:BAM2} vanishes, because for a rigid body of any geometry
\begin{equation}
\intop_{\mathcal{B}_{t}}\boldsymbol{d}\times\rho\left(\boldsymbol{R}_{t}\right)\frac{\hat{\boldsymbol{\xi}_{j}}}{\left|\boldsymbol{\xi}_{j}\right|^{2}}\mathrm{d}V_{t}
=\intop_{\mathcal{B}_{t}}\rho\left(\boldsymbol{R}_{t}\right)\boldsymbol{d}\mathrm{d}V_{t}\times\frac{\boldsymbol{\xi}_{j}}{\left|\boldsymbol{\xi}_{j}\right|^{3}}=0,
\label{eq:BAM2term1}
\end{equation}
given the definition of $\boldsymbol{d}$ as positions of material points in the body from the center of mass.

Noting that the second term in the right hand side of \eqref{eq:BAM2} also vanishes as a result of the cross product of a vector $\boldsymbol{d}$ with itself, we have a single first-order correction in $\varepsilon_{j}$, which simplifies \eqref{eq:BAM} to
\begin{equation}\begin{aligned}
&\intop_{\mathcal{B}_{t}}\frac{\mathrm{d}}{\mathrm{d}t}\left[\left(\boldsymbol{R}_{t}-\boldsymbol{r}_{N}\right)\times\rho\left(\boldsymbol{R}_{t}\right)\boldsymbol{v}_{\mathrm{tan}}\right]\mathrm{d}V_{t}\\
& \qquad \sim 3\sum_{j=1}^{N-1}Gm_{j}\intop_{\mathcal{B}_{t}}\left(\boldsymbol{R}_{t}-\boldsymbol{r}_{N}\right)
\times\rho\left(\boldsymbol{R}_{t}\right)\frac{\left(\left(\boldsymbol{r}_{j}-\boldsymbol{r}_{N}\right)\cdot\left(\boldsymbol{R}_{t}-\boldsymbol{r}_{N}\right)\right)\left(\boldsymbol{r}_{j}-\boldsymbol{r}_{N}\right)}{\left|\boldsymbol{r}_{j}-\boldsymbol{r}_{N}\right|^{5}}\mathrm{d}V_{t},
\label{eq:BAMstep}
\end{aligned}\end{equation}
as $\varepsilon_{j}\rightarrow 0$.

We note that the preceding equations of motion can also be derived from a gravitational potential argument, an example of which is given in \cite{CellettiBook2010} for the case of the Beletsky dumbbell.

\subsection{Reduction to a one-dimensional rotating object in a plane}
\begin{figure}
\begin{centering}
\includegraphics[scale=0.3]{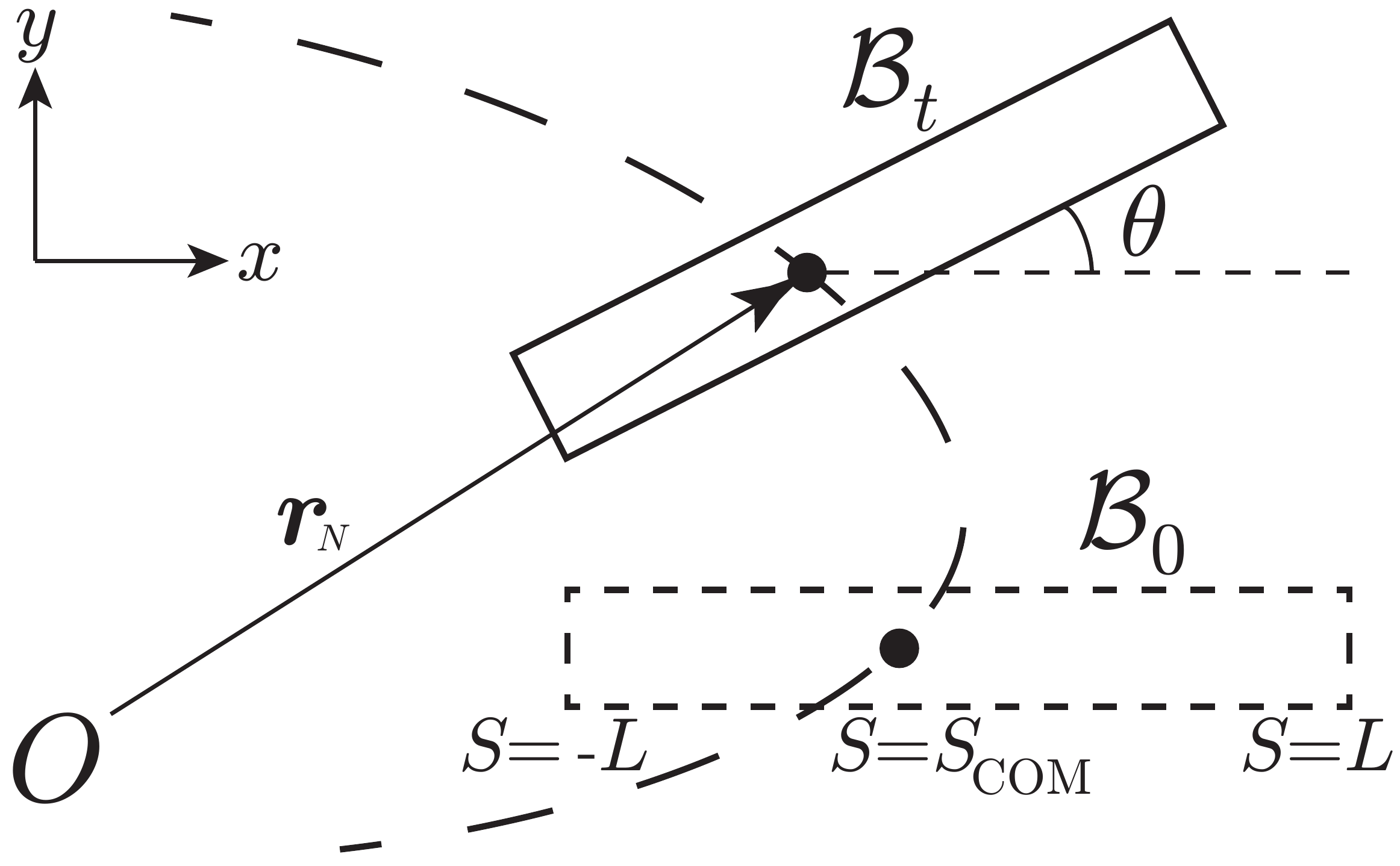}
\par\end{centering}
\caption{The planar motion of a rotating rigid rod is described by a mapping
from a reference configuration $\mathcal{B}_{0}$
to a current configuration $\mathcal{B}_{t}$ at
time $t$. $\mathcal{B}_{0}$ is taken to align with
the positive $x$-axis and is parameterized by a reference arclength
coordinate $S\in\left[-L,L\right]$, whereby the center of
mass occurs at $S=S_{\mathrm{COM}}$ using (\ref{eq:COM}). The position vector $\boldsymbol{r}_{N}$ points
to the center of mass at a particular time, so that the displacement
of a material point in $\boldsymbol{R}_{t}=(x_{N},y_{N}) + \left(S-S_{\mathrm{COM}}\right)\left(\cos\theta,\sin\theta\right)$.\label{fig:RodModel} }
\end{figure}

To obtain an explicit relation for the balance of linear and angular momentum in the quadrupole limit, we suppose that all motion, rotational and translational, is confined to the $x$-$y$ plane and we take the rigid body to have a simple geometry; namely, a one dimensional object. The position of the $i$th center of mass is given by $\left(x_{i}, y_{i}\right)$ in Cartesian coordinates. We describe the rod with an arclength coordinate in the reference configuration $S\in\left[-L,L\right]$ which is assumed to align with the $x$-axis for convenience. Under this reduction, the rigid body rotates around a single axis pointing perpendicular to the plane which we parameterize by an angular position $\theta$ with respect to the positive $x$-axis, so that the position vector of material points in the current configuration is kinematically given by 
\begin{equation}
\boldsymbol{R}_{t}=(x_{N},y_{N}) + \left(S-S_{\mathrm{COM}}\right)\left(\cos\theta,\sin\theta\right)
\end{equation} 
with $\boldsymbol{r}_{N} = (x_N,y_N)$ and where $S_{\mathrm{COM}}$ is the center of mass in this one-dimensional parameterization, defined as (see Fig. \ref{fig:RodModel})
\begin{equation}
S_{\mathrm{COM}}=\frac{1}{m_N}\intop_{-L}^{L}\rho\left(S\right)S\mathrm{d}S.
\label{eq:COM}
\end{equation}
See Fig.~\ref{fig:RodModel} for a diagram of this coordinate system. Note too that we have assumed that the density is parameterized in this geometry as $\rho(S)$.

With these assumptions, we simplify \eqref{eq:BLMPointmassFinal} to obtain
\begin{multline}
\left(\frac{\mathrm{d}^{2}x_{i}}{\mathrm{d}t^{2}},\frac{\mathrm{d}^{2}y_{i}}{\mathrm{d}t^{2}}\right)=\sum_{j\neq i}^{N}\frac{Gm_{j}}{\left[\left(x_{j}-x_{i}\right)^{2}+\left(y_{j}-y_{i}\right)^{2}\right]^{\frac{3}{2}}}\left(x_{j}-x_{i},y_{j}-y_{i}\right)\\
+\frac{G\left(\cos\theta,\sin\theta\right)}{\left[\left(x_{N}-x_{i}\right)^{2}+\left(y_{N}-y_{i}\right)^{2}\right]^{\frac{3}{2}}}\intop_{-L}^{L}\rho\left(S\right)\left(S-S_{\mathrm{COM}}\right)\mathrm{d}S\\
-\frac{3G\left[\left(x_{N}-x_{i}\right)\cos\theta+\left(y_{N}-y_{i}\right)\sin\theta\right]}{\left[\left(x_{N}-x_{i}\right)^{2}+\left(y_{N}-y_{i}\right)^{2}\right]^{\frac{5}{2}}}\left(x_{N}-x_{i}, y_{N}-y_{i}\right)\intop_{-L}^{L}\rho\left(S\right)\left(S-S_{\mathrm{COM}}\right)\mathrm{d}S,
\end{multline}
where we have decomposed the acceleration field and position vectors of all masses into Cartesian components.

However, we note that the integral terms vanish by the definition of \eqref{eq:COM}, which reduces the translational equations of motion to the standard point mass case with higher order corrections occurring at the order of $\varepsilon^{2}_{i}$. We observe that similar reasoning leads to the same reduction for the orbital motion of the rigid body as \eqref{eq:BLMFinal} becomes, under the reduction of the rigid body to a 1D continuum
\begin{multline}
\left(\frac{\mathrm{d}^{2}x_{N}}{\mathrm{d}t^{2}},\frac{\mathrm{d}^{2}y_{N}}{\mathrm{d}t^{2}}\right)=\sum_{j=1}^{N-1}\frac{Gm_{j}}{\left[\left(x_{j}-x_{N}\right)^{2}+\left(y_{j}-y_{N}\right)^{2}\right]^{\frac{3}{2}}}\left(x_{j}-x_{N},y_{j}-y_{N}\right)\\
-\frac{Gm_{j}\left(\cos\theta,\sin\theta\right)}{m_{N}\left[\left(x_{j}-x_{N}\right)^{2}+\left(y_{j}-y_{N}\right)^{2}\right]^{\frac{3}{2}}}\intop_{-L}^{L}\rho\left(S\right)\left(S-S_{\mathrm{COM}}\right)\mathrm{d}S\\
+\frac{3Gm_{j}\left[\left(x_{j}-x_{N}\right)\cos\theta+\left(y_{j}-y_{N}\right)\sin\theta\right]}{m_{N}\left[\left(x_{j}-x_{N}\right)^{2}+\left(y_{j}-y_{N}\right)^{2}\right]^{\frac{5}{2}}}\left(x_{j}-x_{N},y_{j}-y_{N}\right)\intop_{-L}^{L}\rho\left(S\right)\left(S-S_{\mathrm{COM}}\right)\mathrm{d}S,
\end{multline}
with higher order terms of $\varepsilon^{2}_{j}$.

We now consider the reduction of the balance of angular momentum of the rigid body (i.e. \eqref{eq:BAMstep}), which gives
\begin{equation}
\begin{aligned}
&\intop_{-L}^{L}\left(S-S_{\mathrm{COM}}\right)\rho\left(S\right)\frac{\mathrm{d}}{\mathrm{d}t}\left[\left(\cos\theta,\sin\theta,0\right)\times\left(\frac{\mathrm{d}x_{\mathrm{tan}}}{\mathrm{d}t},\frac{\mathrm{d}y_{\mathrm{tan}}}{\mathrm{d}t},0\right)\right]\mathrm{d}S\\
& =3\sum_{j=1}^{N-1}Gm_j\intop_{-L}^{L}\left(S-S_{\mathrm{COM}}\right)^2\rho\left(S\right)\left(\cos\theta,\sin\theta,0\right)\\
&\qquad\qquad\qquad \times \left(\frac{\left((x_j-x_N)\cos\theta+(y_j-y_N)\sin\theta\right)(x_j-x_N,y_j-y_N,0)}{\left((x_j-x_N)^2+(y_j-y_N)^2\right)^{5/2}}\right) \mathrm{d}S,
\end{aligned}
\end{equation}
where we note that the distance between the material point and the center of mass $\left(S-S_{\mathrm{COM}}\right)$ is independent of time for a rigid body.

Simplifying both sides, we obtain
\begin{equation}
\begin{aligned}
& \intop_{-L}^{L}\left(S-S_{\mathrm{COM}}\right)\rho\left(S\right)\frac{\mathrm{d}}{\mathrm{d}t}\left[\frac{\mathrm{d}y_{\mathrm{tan}}}{\mathrm{d}t}\cos\theta-\frac{\mathrm{d}x_{\mathrm{tan}}}{\mathrm{d}t}\sin\theta\right]\mathrm{d}S\\
& =\frac{3}{2}\sum_{j=1}^{N-1}Gm_j\intop_{-L}^{L}\left(S-S_{\mathrm{COM}}\right)^2\rho\left(S\right)\left(
\frac{2\left(x_{j}-x_{N}\right)\left(y_{j}-y_{N}\right)\cos2\theta}{\left(\left(x_{j}-x_{N}\right)^{2}+\left(y_{j}-y_{N}\right)^{2}\right)^{5/2}}\right.\\
&\qquad\qquad\qquad\qquad\qquad\qquad\qquad \left. +\frac{\left[\left(y_{j}-y_{N}\right)^{2}-\left(x_{j}-x_{N}\right)^{2}\right]\sin2\theta}{\left(\left(x_{j}-x_{N}\right)^{2}+\left(y_{j}-y_{N}\right)^{2}\right)^{5/2}}\vphantom{}\right)\mathrm{d}S,
\end{aligned}
\end{equation}
where the left hand term in square brackets is the rotational velocity
multiplied by the distance between the material point and the rod's center of mass, $\left(S-S_{\mathrm{COM}}\right)\frac{\mathrm{d}\theta}{\mathrm{d}t}$. To show this, one could consider the relationship between the angular velocity $\boldsymbol{\omega}$ and the tangential translational velocity $\boldsymbol{v}_{\mathrm{tan}}$
\begin{equation}
\boldsymbol{\omega}=\frac{\left(\boldsymbol{R}_{t}-\boldsymbol{r}_{N}\right)\times\boldsymbol{v}_{\mathrm{tan}}}{\left|\boldsymbol{R}_{t}-\boldsymbol{r}_{N} \right|^2},
\label{RotAcceleration}
\end{equation}
which, for the present case, becomes
\begin{equation}
\frac{\mathrm{d} \theta}{\mathrm{d}t} = \frac{\left( \frac{\mathrm{d}y_{\mathrm{tan}}}{\mathrm{d}t}\cos\theta-\frac{\mathrm{d}x_{\mathrm{tan}}}{\mathrm{d}t}\sin\theta\right)}{(S-S_{\mathrm{COM}})}.
\end{equation}
 
With this, the balance of angular momentum for the rotating rod reduces to
\begin{equation}
\frac{\mathrm{d}^2 \theta}{\mathrm{d}t^2} = \frac{3G}{2}\sum_{j=1}^{N-1}\frac{m_j\left(2\left(x_{j}-x_{N}\right)\left(y_{j}-y_{N}\right)\cos2\theta+\left[\left(y_{j}-y_{N}\right)^{2}-\left(x_{j}-x_{N}\right)^{2}\right]\sin2\theta\right)}{\left(\left(x_{j}-x_{N}\right)^{2}+\left(y_{j}-y_{N}\right)^{2}\right)^{5/2}},
\label{eq:BAMFinal}
\end{equation}
where the integration over $S$, corresponding to the moment of inertia of the rigid body, cancels from both sides.


\end{document}